\documentclass[review]{elsarticle2}
\usepackage{hyperref}
\usepackage{color}

\usepackage{textcomp}

\begin{document}

\begin{frontmatter}

\title{The HPS electromagnetic calorimeter}

\newcommand{\red[1]}{{\color{red}{\bf #1}}}
\newcommand{\JLAB}{Thomas Jefferson National Accelerator Facility, Newport News, Virginia 23606}
\newcommand{\YEREVAN}{Yerevan Physics Institute, 375036 Yerevan, Armenia}
\newcommand{\ODU}{Old Dominion University, Norfolk, Virginia 23529}
\newcommand{\genova}{Istituto Nazionale di Fisica Nucleare, Sezione di Genova e Dipartimento di Fisica dell'Universit\`a, 16146 Genova, Italy}
\newcommand{\SACLAY}{Irfu/SPhN, CEA, Universit\'e Paris-Saclay, 91191 Gif-sur-Yvette, France}
\newcommand{\ORSAY}{Institut de Physique Nucl\'eaire, CNRS/IN2P3, Universit\'e Paris-Saclay, Orsay, France}
\newcommand{\UNH}{University of New Hampshire, Department of Physics, Durham, NH 03824}
\newcommand{\SLAC}{SLAC National Accelerator Laboratory, Menlo Park, CA 94025}
\newcommand{\GLASGOW}{University of Glasgow, Glasgow, G12 8QQ, Scotland, UK}
\newcommand{\ROMA}{Istituto Nazionale di Fisica Nucleare, Sezione di Roma Tor Vergata and Dipartimento di Fisica dell'Universit\`a di Roma Tor Vergata, 00133 Roma, Italy}
\newcommand{\TORINO}{Istituto Nazionale di Fisica Nucleare Sezione di Torino, 10125 Torino, Italy}
\newcommand{\CATANIA}{Istituto Nazionale di Fisica Nucleare, Sezione di CATANIA 95123, Catania, Italy}
\newcommand{\CATA}{Istituto Nazionale di Fisica Nucleare, Laboratori Nazionali del Sud and University of Sassari, 95123, Catania, Italy}

\author[TORINO]{I.~Balossino}
\author[JLAB]{N.~Baltzell} 
\author[GENOVA]{M.~Battaglieri}
\author[CATANIA]{M.~Bond\`i} 
\author[GLASGOW]{E.~Buchanan}
\author[TORINO]{D.~Calvo}
\author[GENOVA]{A.~Celentano} 
\author[ORSAY]{G.~Charles\corref{corrauthor}}
\ead{charlesg@ipno.in2p3.fr}
\author[ORSAY,ROMA]{L.~Colaneri}
\author[ROMA]{A.~D'Angelo}
\author[CATANIA]{M.~De~Napoli} 
\author[GENOVA]{R.~De~Vita}
\author[ORSAY]{R.~Dupr\'e}
\author[JLAB]{H.~Egiyan}
\author[ODU]{M.~Ehrhart}
\author[TORINO]{A.~Filippi}
\author[JLAB,SACLAY]{M.~Gar\c con\corref{corrauthor}}
\ead{michel.garcon@cea.fr} 
\author[YEREVAN]{N.~Gevorgyan} 
\author[JLAB]{F.-X.~Girod} 
\author[ORSAY]{M.~Guidal}
\author[UNH]{M.~Holtrop}      
\author[ORSAY]{V.~Iurasov}
\author[JLAB]{V.~Kubarovsky}
\author[GLASGOW]{K.~Livingston} 
\author[UNH]{K.~McCarty}
\author[SLAC]{J.~McCormick}
\author[GLASGOW]{B.~McKinnon}
\author[GENOVA]{M.~Osipenko} 
\author[UNH]{R.~Paremuzyan}
\author[CATANIA]{N.~Randazzo}
\author[ORSAY]{E.~Rauly}
 \author[JLAB]{B.~Raydo} 
\author[ORSAY]{E.~Rindel}
\author[ROMA]{A.~Rizzo}
\author[ORSAY]{P.~Rosier}
\author[CATA]{V.~Sipala}
\author[JLAB]{S.~Stepanyan}
\author[ODU]{H.~Szumila-Vance}
\author[ODU]{L.~B.~Weinstein}        
%
%
%
%
%
%
 %
%
%
%
%
%
%
%
%

\address[TORINO]{\TORINO}
\address[JLAB]{\JLAB}
\address[GENOVA]{\genova}
\address[CATANIA]{\CATANIA}
\address[GLASGOW]{\GLASGOW}
\address[ORSAY]{\ORSAY}
\address[ROMA]{\ROMA}
\address[ODU]{\ODU}
\address[SACLAY]{\SACLAY}
\address[YEREVAN]{\YEREVAN}
\address[UNH]{\UNH}
\address[SLAC]{\SLAC} 
\address[CATA]{\CATA}


\cortext[corrauthor]{Corresponding authors:}

\begin{abstract}
The Heavy Photon Search experiment (HPS) is searching for a new gauge boson,
the so-called ``heavy photon." Through its kinetic mixing with the
Standard Model photon, this particle could decay into an electron-positron pair. It would then
be detectable as a narrow peak in the invariant mass spectrum of such pairs,
or, depending on its lifetime, by a decay downstream of the production target.
The HPS experiment is installed in Hall-B of Jefferson Lab.
This article presents the design and performance of one of the two detectors of the
experiment, the electromagnetic calorimeter, during the runs performed in 2015-2016.
The calorimeter's main purpose is to provide a fast trigger and reduce the copious background from electromagnetic processes
through matching with a tracking detector. The detector is a homogeneous calorimeter,
made of 442 lead-tungstate (PbWO$_4$) scintillating crystals, each read out by an avalanche photodiode
coupled to a custom trans-impedance amplifier.

\end{abstract}


\end{frontmatter}

\section{Introduction}
\begin{figure}[htb]
  \begin{center}
     \includegraphics[angle=0, width=0.9\textwidth]{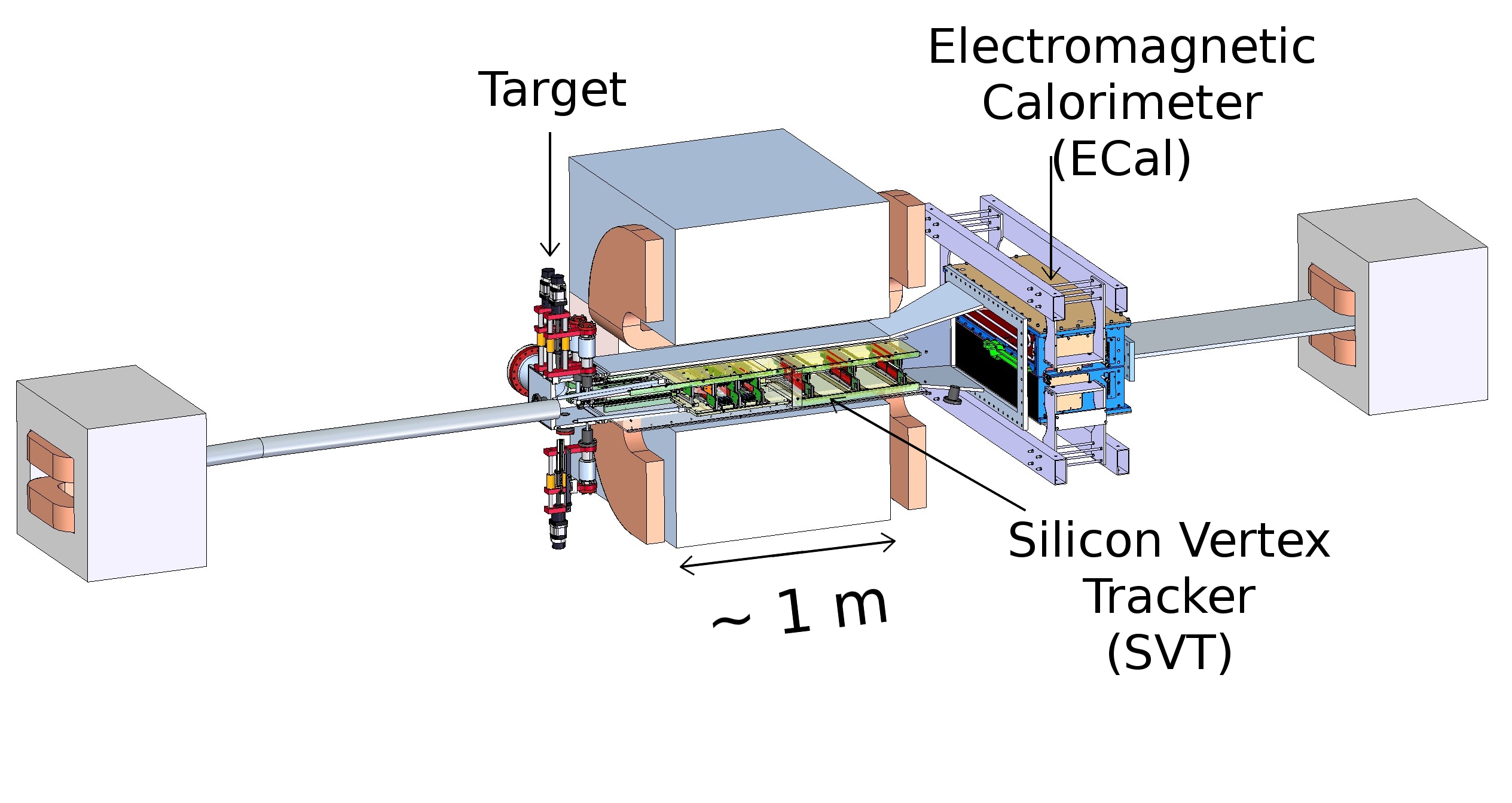}
    \caption{Schematic layout of the HPS experiment: the beam coming from the left is deflected
     toward the target placed at the entrance of the second analyzing dipole magnet. The SVT is located inside the gap of this dipole,
     and the ECal right after. The last magnet steers the beam back into the initial direction towards the beam dump.
	}
    \label{intro:HPSScheme}
  \end{center}
\end{figure}

The heavy photon, also known as $A'$ or dark photon, is a conjectured massive gauge boson
associated with a new U(1) hidden symmetry, and a possible force carrier
between dark matter particles.
Such a heavy photon has been envisioned by several theories beyond the Standard Model and
is also a good candidate to explain some existing astrophysical anomalies.
The $A'$ would interact with particles of the hidden sector
and kinetically mix with the ordinary photon~\cite{REssig}.
This kinetic mixing generates its weak
coupling to electrons allowing heavy photons to be radiated in electron
scattering and subsequently decay into electron-positron pairs. If the
coupling is large enough, the decay products should be observable above the QED
background in the $e^+e^-$ invariant mass spectrum, while if the coupling is small, heavy photons would travel
detectable distances before decaying. The HPS experiment is designed to
exploit both signatures. Benefitting from the full duty cycle of the
electron beam available at Jefferson Lab, several data-taking runs have been taken and more are planned with beam
energies between 1~GeV and 6.6~GeV. The electron beam, of intensity between
50~nA and 400~nA, impinges on 0.15\%~-~0.25\% radiation length tungsten foils.
A silicon microstrip vertex tracker (SVT) begins 10~cm downstream of the target within the gap of a dipole magnet for the determination of the leptons' momenta and angles, while the
electromagnetic calorimeter (ECal) is located 139~cm from the target, outside that magnet,
and serves primarily as a fast trigger.
Both SVT and ECal are placed as close as possible to the horizontal plane containing the beam, thus allowing
the detection of lepton pairs with very small opening angles and providing sensitivity
to heavy photons in the mass range from 20~MeV/c$^2$ to
1~GeV/c$^2$.
The experiment is installed in Hall-B at Jefferson Lab, positioned
at the downstream end of the hall
in the configuration illustrated in Fig.~\ref{intro:HPSScheme}.
\nopagebreak

A test run was performed in May 2012 with a partial detector setup, as described
in Ref.~\cite{HPSTestRun}. We detail here the design and performance of the final calorimeter configuration used during the 2015-2016 engineering runs.
 The paper is organized as follows.
 We first present the calorimeter design and layout, with special emphasis on the modifications made after the test run.
The next Section deals with Monte Carlo simulations of the detector response.
The ECal performance, obtained after time and energy calibrations, is discussed in Section 4.
Section 5 is devoted to the trigger performance.
Section 6 addresses some aspects of SVT track and ECal cluster matching and
is followed by a summary of calorimeter's impact on the experiment.

\section{Calorimeter description}
In order to provide a reliable trigger in a high-background environment (up to 1~MHz/cm$^2$),
the HPS electromagnetic calorimeter must be fast and match the lepton-pair acceptance of the SVT, while operating in the fringe field of the analyzing dipole.
Electrons and positrons between 0.3~GeV and 6.6~GeV are to be measured
with an energy resolution of the order of 4\%/$\sqrt{E}$ and a position resolution of about 1 to 2~mm to match with the corresponding information coming from the SVT.

For these purposes, a homogeneous calorimeter made of lead tungstate (PbWO$_4$) scintillating crystals was constructed. PbWO$_4$ crystals have a fast decay time ($\approx$ 10~ns) thus allowing operations in the HPS high-rate environment with a reduced pile-up probability, and a reasonable light yield, compatible with the energy resolution requirements for this experiment. The calorimeter was constructed with refurbished crystals originally used in the inner calorimeter of the CLAS detector~\cite{FX-thesis}. Given the presence of magnetic fringe fields, avalanche photodiodes were used for light readout, coupled to custom preamplifiers.
Major improvements with respect to the test-run configuration included larger-area avalanche photodiodes, optimized low-noise preamplifiers, new mother-boards for the routing of high-voltage and signals, and a new light-monitoring system.

\subsection{Crystals and ECal lay-out}
\begin{figure}[bht]
\centering
\includegraphics[width=1.00\textwidth]{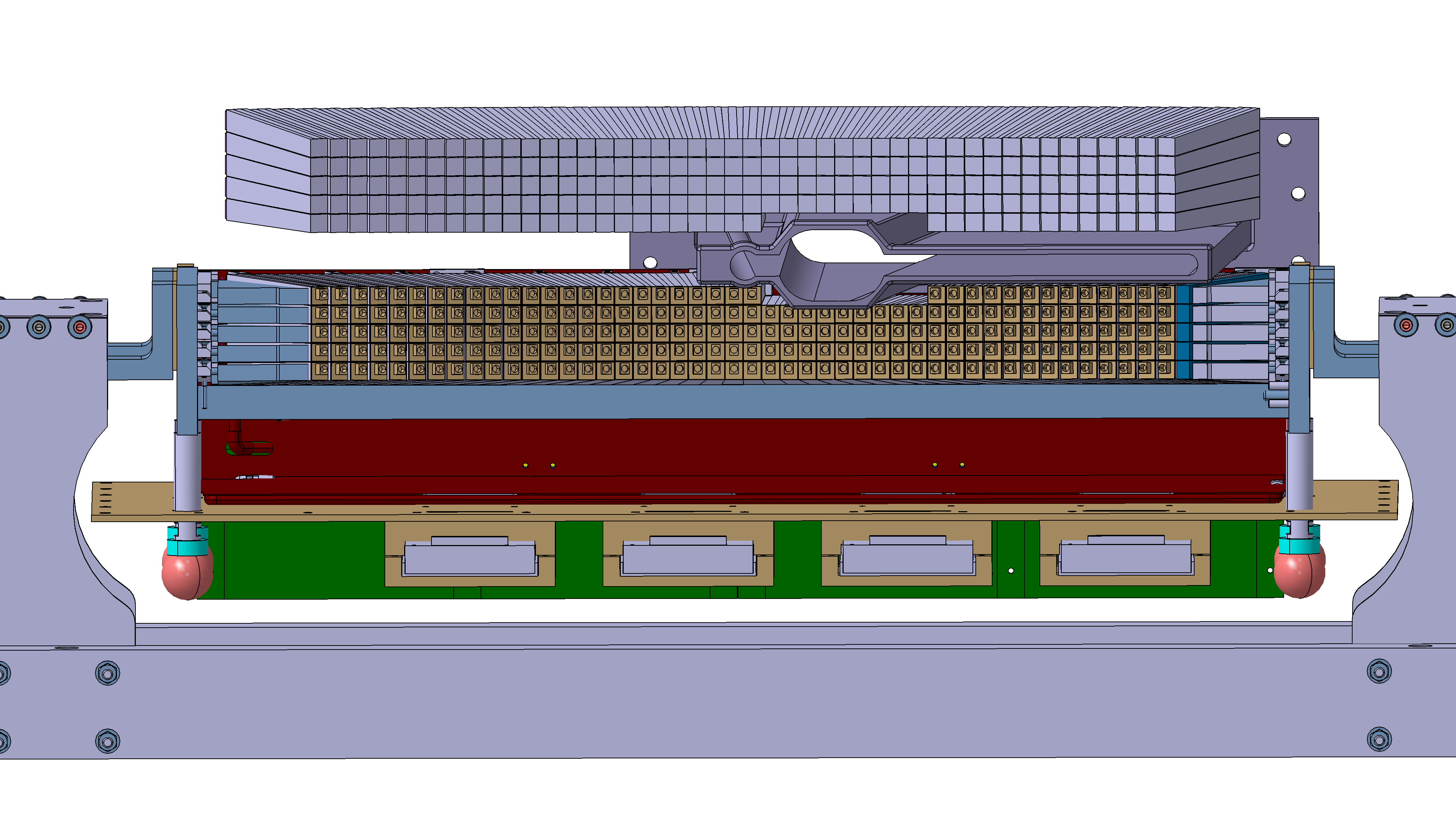}
\caption{ECal crystal layout, as seen in the beam direction. For clarity, the top-half mechanical parts have been removed. For the bottom half, some mechanical elements such as the mother boards (in green) and the copper plates for heat shielding (in red) are visible.
		Between the two halves of ECal, the beam vacuum vessel is seen to be extended to the right to accommodate
		beam particles having lost energy through scattering or radiation.}
\label{Calo}
\end{figure}

The calorimeter design is shown in Fig.~\ref{Calo}. In order to avoid
a vertical 15~mrad zone of excessive electromagnetic background,
 the ECal is built as two separate halves that are mirror reflections of
one another around the horizontal plane.

\begin{figure}[ht!]
\centering
\includegraphics[width=0.70\textwidth]{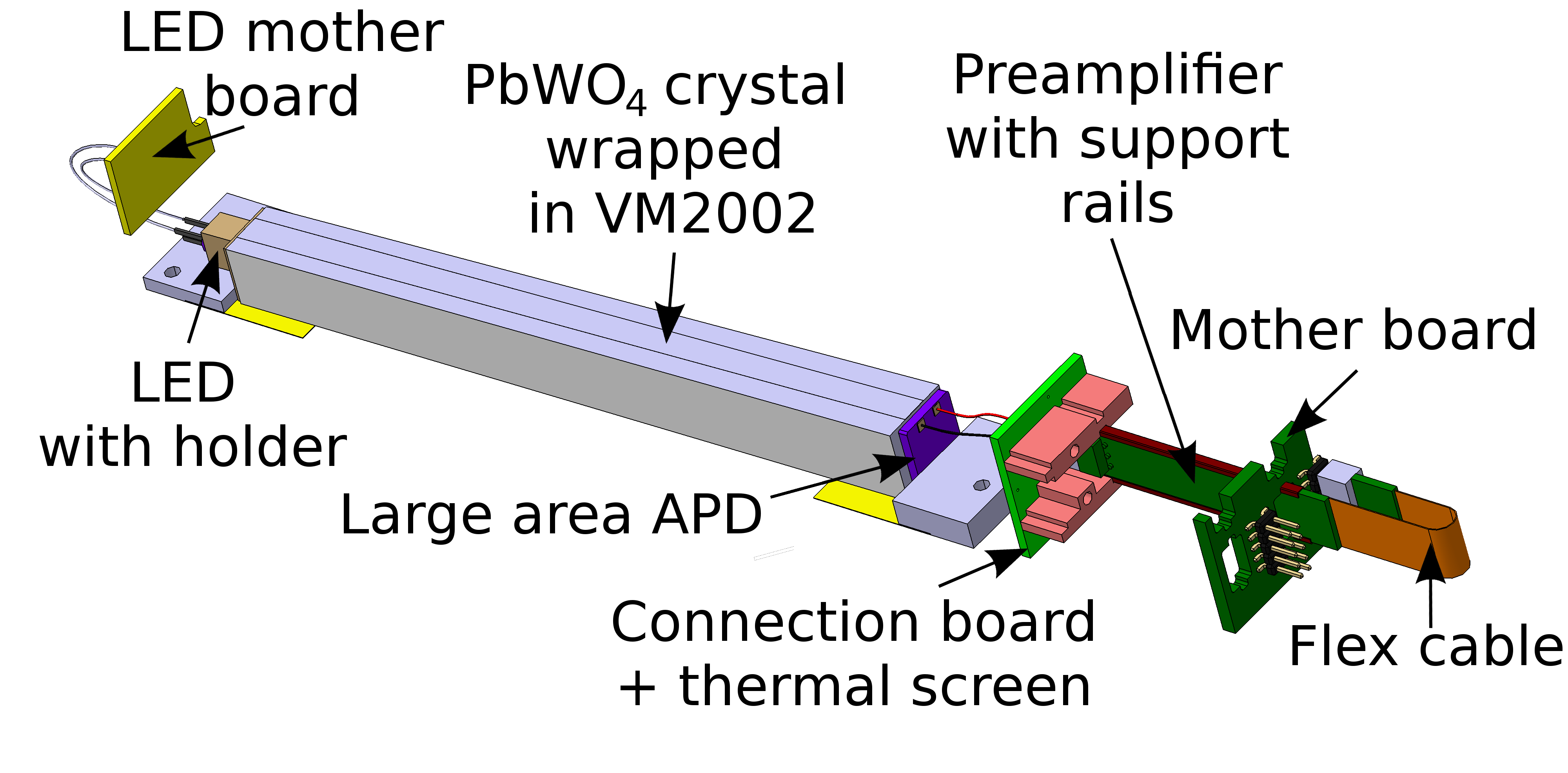}
\caption{A schematic view of an ECal module.}
\label{Crys}
\end{figure}

Each half is made of 221 modules supported by aluminum frames and arranged in a rectangular structure with five layers of 46 crystals.
The two layers closest to the beam have 9 modules removed to allow a larger opening for the outgoing, partially degraded,
electron beam and copious Bremsstrahlung photons. The layout of a single module is shown in Fig.~\ref{Crys}.

Each module is made of a 160-mm long (18 radiation lengths), tapered PbWO$_4$ crystal, with a front (rear) face of 13.3$\times$13.3~mm$^ {2}$ (16$\times$16~mm$^{2}$), wrapped in a VM2002 reflecting foil to increase light collection.
The 10$\times$10~mm$^ {2}$ Hamamatsu photo-sensor~\cite{APD} is glued to the rear face and connected to the preamplifier~\cite{preampli} held in position by a connection board that also serves as a thermal screen.
The signal is then routed through a mother board to the external DAQ system.
On the front face, each crystal hosts a bi-color light-emitting diode (LED) that serves as a monitoring device, as described later.

To stabilize the crystal light yield and the APD gains, each half of the calorimeter is enclosed
in a temperature-controlled box. A chiller, operating at 17$^{\circ}$C, circulates water through these
enclosures and maintains a stability to better than $0.3^{\circ}$C.

Both halves were held in place by vertical threaded rods
attached to rails above the analyzing magnet.
The gap between the two halves was determined to be 44~mm,
very close to the design value of 40~mm,
reproducible to within 0.3~mm after displacing them vertically in order to perform maintenance work on the SVT, the vacuum system or the ECal itself.


\subsection{Light detection and electronics}
The major upgrade of the calorimeter system, compared to the 2012 test-run configuration, is the introduction of new 10$\times$10~mm$^2$ Large-Area APDs from Hamamatsu. At equivalent deposited energy, about four times more light is collected compared to the 5$\times$5~mm$^2$ APDs used for the test run.
The signal-over-noise ratio is thus increased, allowing for a lower energy threshold and an improved energy resolution.

\begin{figure}[ht!]
\centering
\includegraphics[width=0.90\textwidth]{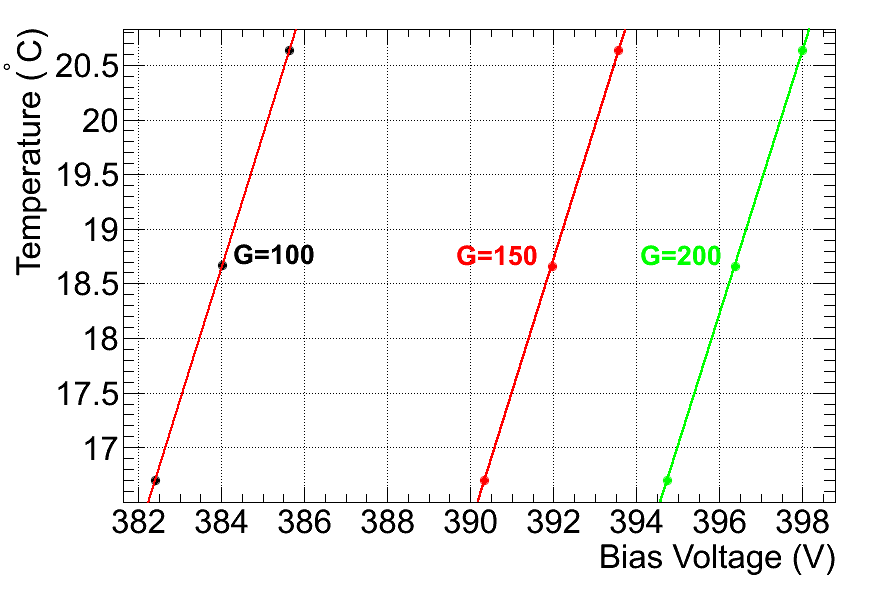}
\caption{Gain variation with bias voltage and temperature for a typical APD.
Circles are interpolations between measurements at fixed temperature and varying voltage.
Lines are iso-gain curves in the temperature - bias voltage plane. }
\label{TGV}
\end{figure}

The dependence of the gain and leakage current on the bias voltage and temperature were measured in a specially designed test bench~\cite{LEDbench}.
A typical result is illustrated in Fig.~\ref{TGV}, where the gain is seen to depend on voltage and temperature through a linear combination: $G=G(\alpha V - \beta T)$.
After characterizing each photo-sensor, the operating voltage was chosen to yield the best compromise between a high gain and a low dark current.
The APDs were then grouped into ensembles of 4 to 10 with similar
gain-to-voltage characteristics so that each group could be powered by a
single high-voltage channel.
The bias voltage of each APD group was selected to ensure an average gain of 150 at 18$^\circ$C for each APD.

The signal from the APD is sent to a preamplifier
which converts current to voltage and has low
input impedance and noise.
The gain of the preamplifier was adjusted to
ensure that the maximum energy deposition, estimated to be 4~GeV in a single
crystal for a beam energy of 6.6 GeV, would not saturate the amplitude converter.

Typical numbers characterizing the whole chain crystal-APD-preamplifier are~\cite{Andrea-Gabriel-Note}:
a light yield of 120 photons (reaching the rear face of the crystal) per MeV of deposited energy,
an APD quantum efficiency of 0.7 and gain of 150, a preamplifier gain of 0.62~V/pC (for a 10~ns input pulse width) and a maximal signal amplitude of 2~V.
A noise level of a few mV allows thresholds to be placed on individual crystals which are equivalent to 7.5~MeV.

The APD  bias voltages (between 385~V and 405~V), the operating voltage of the preamplifiers ($\pm$5~V),
 and their ouput signals are distributed through four circuit boards, known as mother boards.
These were completely redesigned after the 2012
test run and careful attention was paid to avoid cross-talk
between channels. Each half of the ECal is divided into 26 bias voltage
groups. By properly selecting these groups, matching them with appropriate preamplifiers, and fine-tuning the high-voltages, a total gain uniformity of the order of a few percent was achieved. Up to fluctuations in the light-yield from crystal to crystal,
this provided a good starting point of operation for the trigger set-up.

Finally, for digitization and processing,
the signal  was sent to a Jefferson Lab Flash ADC (FADC250) board~\cite{HPSTestRun, FADC}.
The FADC digitizes the APD signal at 250 MHz and stores samples in an 8 $\mu$s deep pipeline with 12-bit resolution.

\subsection{Slow controls}

The slow controls and monitoring of the calorimeter systems are all implemented within the
EPICS framework~\cite{EPICS}.  Graphical interfaces are used for easy user interaction, as well as an alarm
system with audible alerts in the control room and automatic email alerts to system experts.
Time histories of all slow controls data are preserved and accessed with Jefferson Lab's MYA archiving system \cite{mya}.

The ECal water cooling is provided by an Anova A-40 chiller operating at 17$^{\circ}$C.
The internal temperature of the calorimeter is also monitored using sixteen
thermocouples located throughout the crystal lattice. The thermocouples are read-out using
Omega D5000 series transmitters. Both devices provide RS-232 serial communications.

Low voltage is supplied to the preamplifiers via an Agilent 6221 running at $\pm$~5~V and 4~A and
 remotely controlled and monitored from EPICS through a GPIB-ethernet converter.

The serial communications with the Anova and Omega devices are all via
a MOXA N-Port 5650 serial to ethernet device server and shielded 50-foot cables
between the detector and electronics areas.  These are then interfaced with EPICS
via its asynchronous driver for controls and monitoring.

Scalers from FADC modules are read into EPICS from a JLab TCP server running on
their VXS crates.  The current setup provides a graphical display of the 442 scaler channels and a
sampling rate of 1 Hz, well below the limits of both hardware and software.

High voltage is supplied to each of the 52~APD groups via CAEN A1520P modules
in an SY4527 mainframe. Communications with EPICS is achieved via the manufacturer supplied
driver and ethernet connection to the mainframe.

\subsection{LED monitoring}
Although relatively radiation tolerant, lead-tungstate scintillating crystals are subject to a decrease in light output when exposed to radiation. They recover when the radiation source is removed, through spontaneous thermal-annealing mechanisms (see for example Ref.~\cite{Batarin}).
In order to preserve the intrinsic energy resolution, the response of the crystals has to  be continuously monitored and, if necessary, recalibrated. An LED based monitoring system was specifically designed and installed in the detector setup after the 2012 test run.
Glued on the front face of each crystal, a plastic holder hosts a bi-color LED. These LEDs are connected through twisted-pair wires to four printed-circuit boards which are connected to eight driver circuits, externally mounted on top and bottom of the detector enclosure. A red or blue light pulse with variable amplitude and width can be injected independently into each crystal.
By measuring the response of the whole chain (crystal + APD + amplifier) to the pulse, variations in the channel response can be determined and, if necessary, corrected. Furthermore, the radiation damage in the PbWO$_4$ crystals is not uniform over the transmission spectrum as it is mostly concentrated in the blue region (up to $\simeq$ 500 nm). The use of a red/blue bi-color LED can also help in determining which component of the read-out chain is responsible for response variations. During the ECal and trigger commissioning, the LED system was extensively used by turning on one or more channels at a time, sometimes following a programmable pattern.



\section{Simulations}
\begin{figure}[bht]
  \centering
      \includegraphics[width=0.7\textwidth]{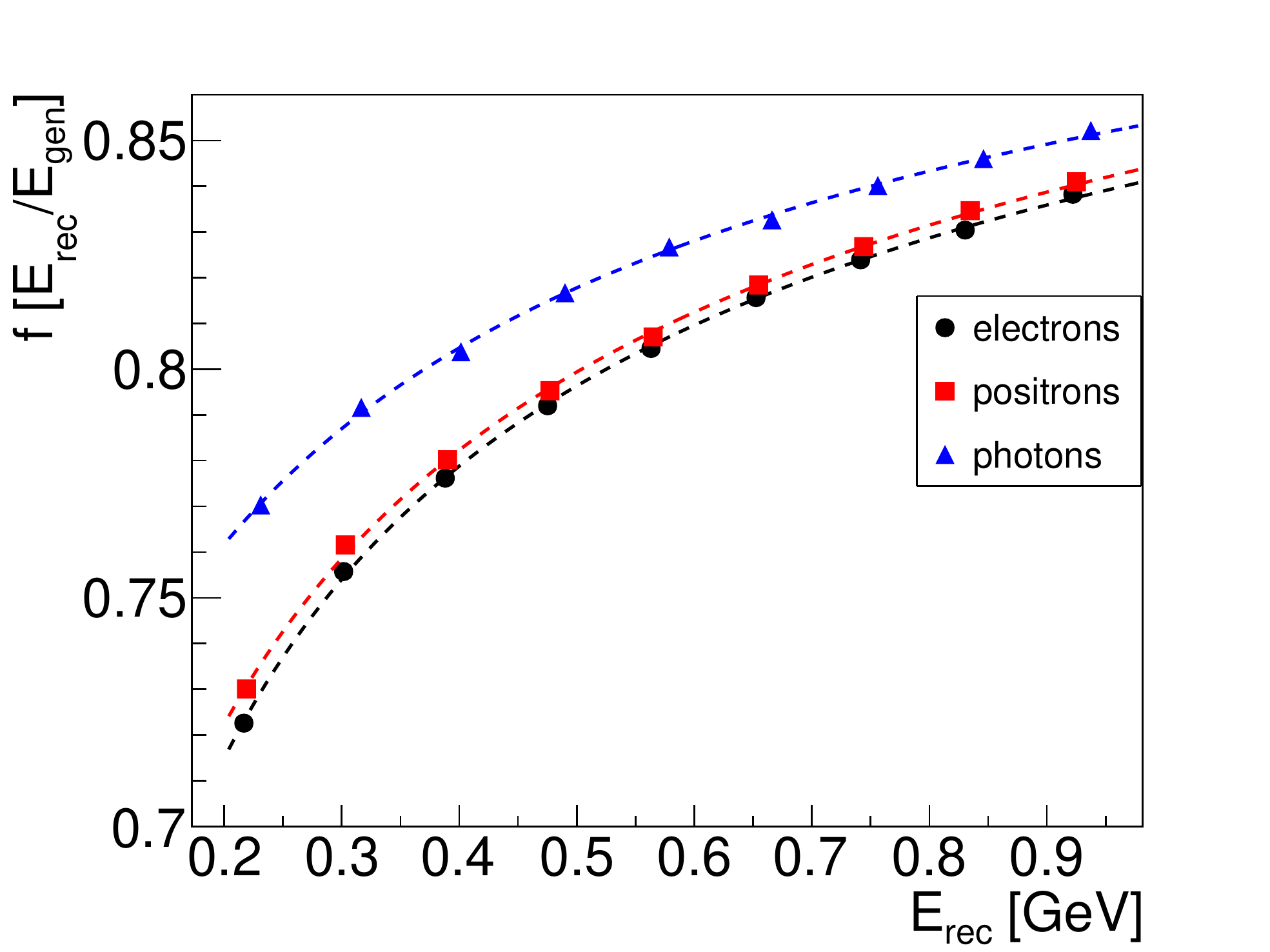}
  \caption{$f$ ratio for electrons, positrons and photons as a function of cluster energy (simulation within a fiducial cut).}
  \label{mcsf}
\end{figure}

A detailed simulation of the electromagnetic showers in the ECal was performed with GEANT4 software \cite{Geant4}
to determine the expected detector performance in terms of energy and position resolutions.
A main goal of the simulations was to calculate the ratio $f$ of the measured cluster energy $E_{rec}$ to the true (generated) energy $E$, as a function of the impinging particle type, energy and position. This ratio $f$ will be referred to as the ``energy correction function" since it is the correction to be applied to the measured energy in order to recover the true energy. Electrons, positrons and photons were simulated at discrete energies, in steps of 0.1 GeV between 0.3 and 1.1 GeV, in order to uniformly cover the range of energies detectable in the run performed at 1.05 GeV beam energy. The same cluster reconstruction code as that used for real data was then applied, and the obtained reconstructed energy was compared to the real value to evaluate $f$. The following thresholds were applied on the measured energy: 7.5 MeV for individual hits (per crystal), 50 MeV for the seed hit in a cluster and 100 MeV for the cluster energy. A seed hit is defined as the crystal with the greatest energy deposition in a given cluster.
Results for $f$ are illustrated in Fig.~\ref{mcsf} for particles hitting the ECal in the fiducial zone, defined as the area occupied by the inner crystals (that is, excluding crystals at the calorimeter edges).
The difference of response between photons and electrons is due in part to the material between the target and the crystals, which is equivalent to 0.47 of a radiation length. This material includes the vacuum exit aluminum window (the dominant contribution), the ECal copper thermal shield and aluminum structure, as well as the SVT. The form of the energy correction function is well described by a 3-parameter fit:
\begin{equation}
\label{eq:samplingfraction}
f \equiv \frac{E_{rec}}{E} = \frac{A}{E_{rec}}+\frac{B}{\sqrt{E_{rec}}}+C .
\end{equation}
	
The incident angle (with respect to the crystal axis) of particles varies with position across the calorimeter. For photons, this is due to the fact that the tapered crystals are pointing downstream of the target position. For electrons and positrons, the deflection in the magnetic field induces energy-dependent impact positions and incident angles. These effects are the other cause of the differences between the energy correction functions for the three particle types seen in Fig.~\ref{mcsf}.

\begin{figure}
  \begin{center}
    \includegraphics[width=0.90\textwidth]{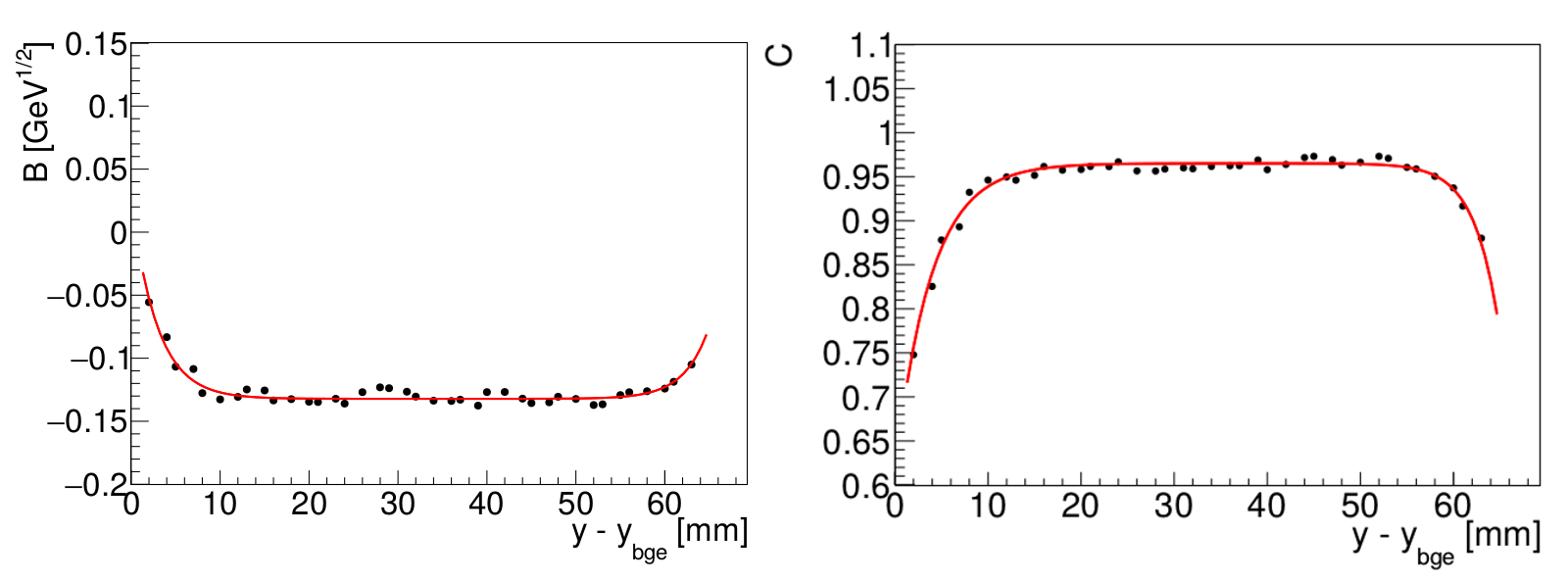}
    \caption{\label{p1_em} Parameters $B$ and $C$ from Eq~\ref{eq:samplingfraction} for electrons, as a function of vertical position relative to the innermost beam gap edge.}
  \end{center}
\end{figure}

The shower leakage effects in the ECal are more important close to its edges. The correction function $f$ was studied and parameterized as a function of distance to the edge of the calorimeter. It is effectively constant in the central region of the ECal but drops off rapidly in the outermost crystals. In Eq.~\ref{eq:samplingfraction}, the parameter $A$ is not significantly correlated with position and remains constant for a given particle type. Moreover, the contribution of $A/E_{rec}$ is small compared to the two other terms. The parameters $B$ and $C$ are strongly correlated with the position of the cluster relative to the ECal edge. This is illustrated in Fig.~\ref{p1_em} for electrons. The true (generated) value is used to determine the position of the particle at the face of the ECal, whereas in data, this position value comes from the SVT tracking.
The parameters $B$ and $C$ are fit with two exponential functions at the edges that match in the central region of the ECal. This procedure was refined to take into account the exact geometry around the beam gap, where there are four crystals in each half-column instead of five.
 Finally, it was extended to the vertical edges with a dependence on the horizontal coordinate. The procedure was repeated for positrons and photons, with the same functional forms but slightly different parameters obtained from the fits.


We point out that these simulations are critical for understanding the electromagnetic shower leakage near the edges of the calorimeter: while the correction function $f$ can be studied with data within the fiducial zone, the full energy correction at the edges is difficult to extract from real data as the energy resolution deteriorates dramatically (see Section~\ref{sec:edges}). The simulated correction function $f$ is used in the real event reconstruction.

The simulation was also used to optimize the determination of the cluster position. The horizontal position $x$ of a cluster was determined by weighting the corresponding crystal centers, $x_i$, with a proper energy-dependent factor $w_i$:
\begin{equation}
x=\frac{\sum_iw_ix_i}{\sum_iw_i} + \Delta x,\ \hbox{with}\ w_i=\max\left[0,w_0+\ln\frac{E_i}{E_{rec}}\right],
\label{eq:xcl}
\end{equation}
and similarly for the vertical position $y$~\cite{Loo}. The parameter $w_0 = 3.1$ acts as a relative energy threshold $E_i/E_{rec} > e^{-w_0}$, while the logarithmic weights favor the lateral tails of the shower for a more precise position determination. In addition, for the horizontal coordinate only, a linear correction $\Delta x(x)$ is added, due to the angle of incidence of the tracks upon the crystal~\cite{SFraction}. This correction depends on the type of particle. These studies showed that the expected resolution on both coordinates is of the order of 2~mm for 1~GeV particles. Considerations on measuring the position using data can be found in Section~\ref{sec:SVTmatching}.

\section{Calorimeter performance}

\subsection{Energy calibration}
\label{EnerCalibSubsection}
Three physical processes were used to calibrate the ECal energy response.
The initial gain calibration was accomplished by measuring the energy deposition from cosmic rays.
The gain coefficients were then refined by using elastically-scattered electrons carrying nearly the full beam energy.
These two calibration points cover the smallest and the largest energies to be measured by the ECal.
Finally, wide-angle Bremsstrahlung events were studied to adjust the simulated correction functions $f$ for mid-range particle energies.
We note that, for optimal energy determination, the energy deposited in each crystal was extracted from a fit to the pulse shape as described in Section~\ref{sec:PulseFitting}.

%

\subsubsection{Calibration with cosmic rays}
\begin{figure}[hbt]
  \centering
      \includegraphics[width=0.65\textwidth]{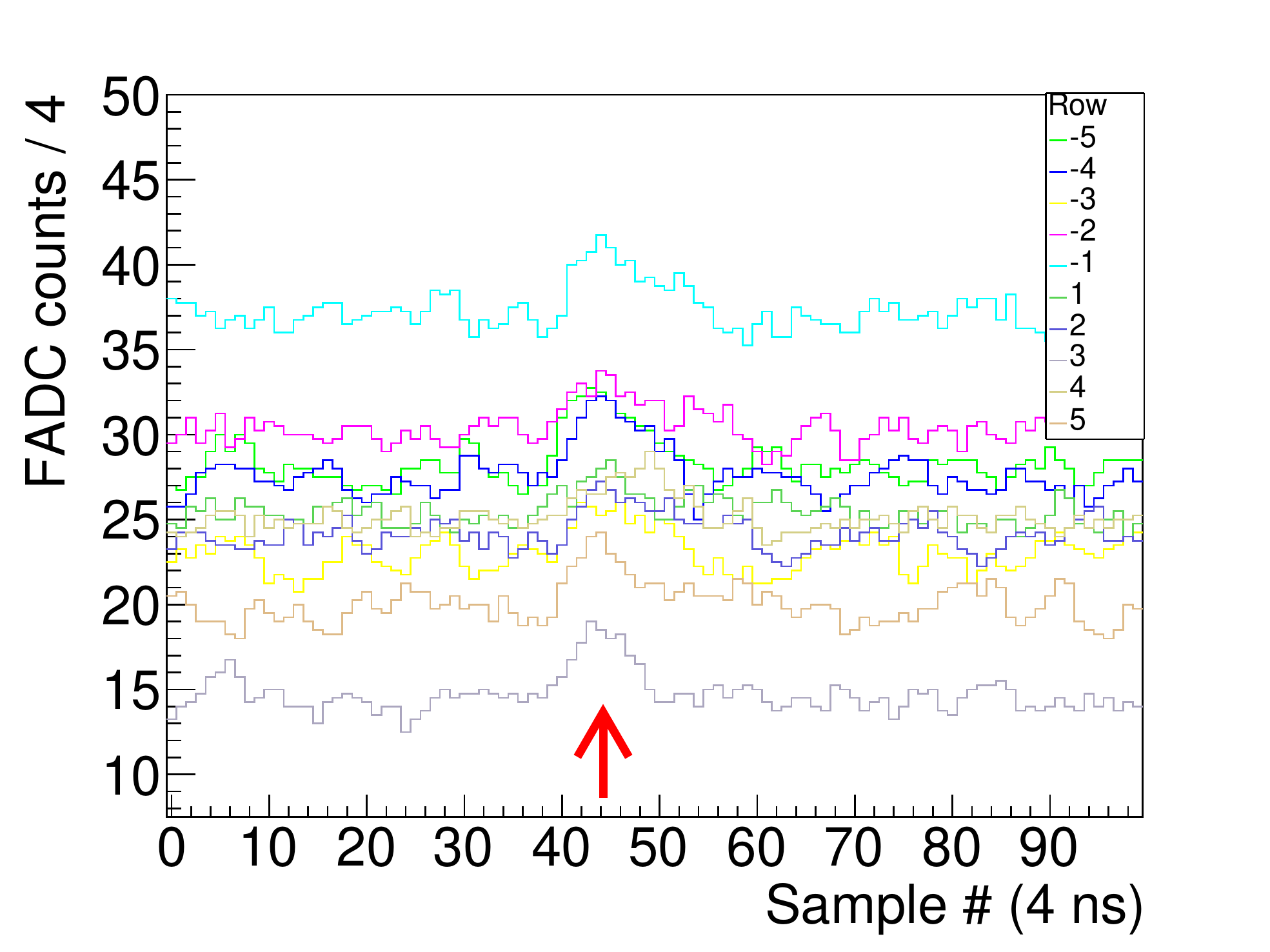}
  \caption{Cosmic ray signal passing vertically through all ten layers of crystals in the ECal. In this plot, each crystal signal is separated vertically  by an offset.}
  \label{cosmicSignal}
\end{figure}

In order to measure the ECal response of nearly vertical cosmic rays, two plastic scintillator paddles were placed below the detector, triggering the readout of all crystals during periods with no beam on target.
Among the tracks collected, only the most vertical ones were kept to minimize the variations of path length across each crystal. This ensured that the energy deposited in the crystal was on average about 18.3~MeV as calculated from simulation.
As an example, the signals from a cosmic ray muon passing vertically through 10 crystals of the ECal can be seen in Fig.~\ref{cosmicSignal}.
As seen from this figure, the signals are close to threshold, but still usable for an initial calibration. This was not the case with the old APDs used during the test run.
A typical cosmic-ray run lasted approximately 60 hours. The integrated charge distributions from all crystals were then fitted to the simulated expectations, and the initial gain calibration was obtained for all 442 channels.
The relative energy resolution obtained after this calibration was around 8\%$/\sqrt{E({\textrm{GeV}})}$. This method for obtaining the gain value of each channel was sufficiently precise for use in the trigger during runs with beam (see Section~\ref{sec:Trigger}).

\subsubsection{High-energy calibration with elastically-scattered electrons}
The energy of electrons elastically scattered from the target through small angles which are detected in the ECal peaks at the beam energy (within 0.1\%).
For this calibration, only clusters for which the seed hit carried more than 60\% of the full cluster energy were kept.
Furthermore, the seed hit energy was required to be larger than 450 MeV during the run at 1.05~GeV beam energy, and larger than 1.1~GeV during the run at 2.3~GeV.
A given crystal was calibrated using all clusters for which it is the seed.
The high-energy calibration resulted from the comparison of the measured cluster energy with the one expected from simulations.
Since the procedure involves using the full cluster energy, thus including information from multiple crystals,
it was iterated until all values of individual crystal gains were stable to within 1\%.	
Two iterations of the procedure were sufficient to reach the calibration of the 366 crystals that had geometric acceptance for
elastically-scattered electrons. Due to the combined effect of geometry and magnetic field deflection, elastically-scattered electrons cannot reach several crystals on both right and left sides of the ECal.

\begin{figure}[hbt]
   \centering
	\includegraphics[height=0.65\textwidth]{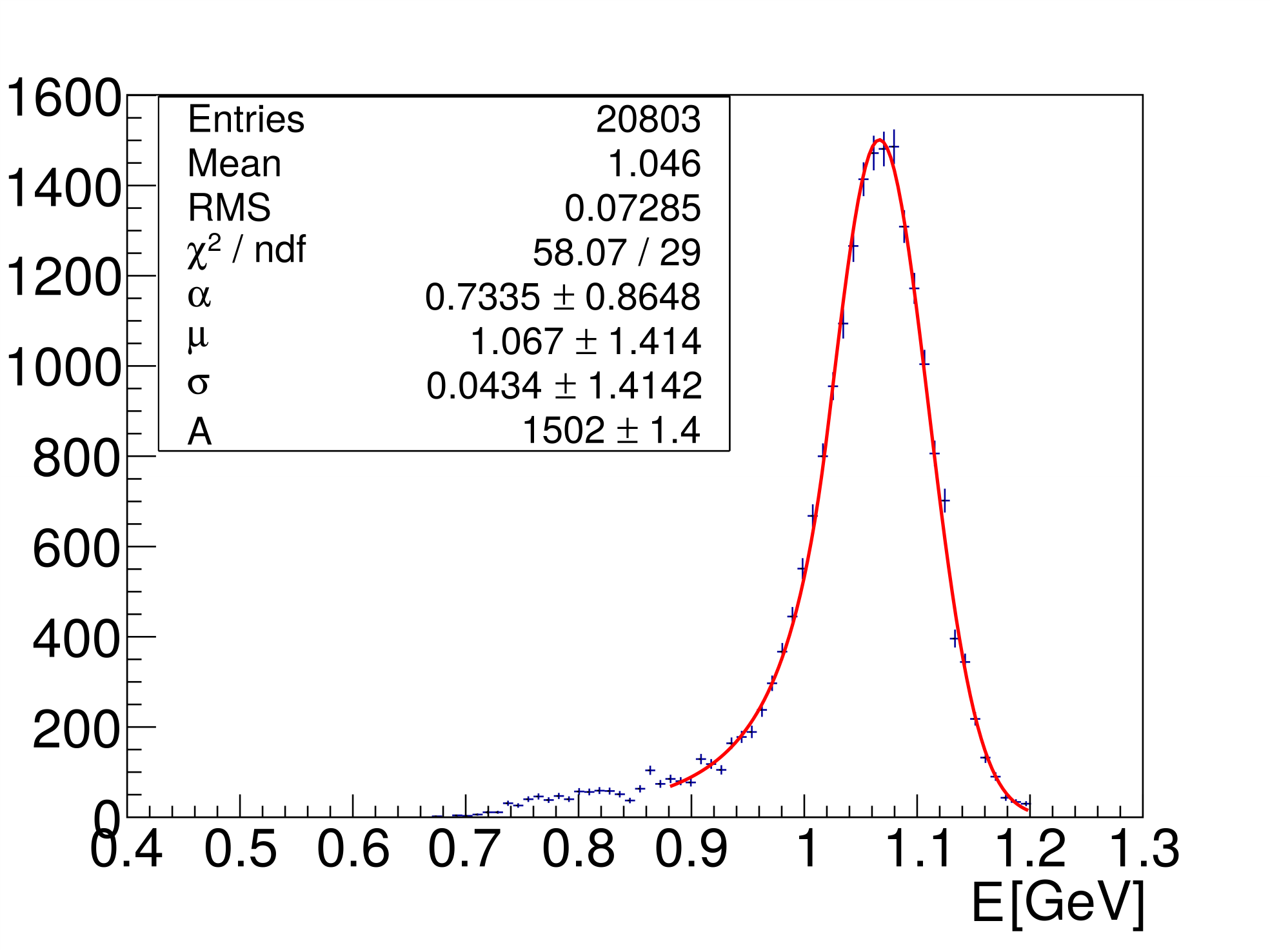}
	\caption{Measured energy of elastically-scattered electrons after calibration and correction $f$ due to shower leakage, at a beam energy of 1.05~GeV.
	 This plot sums over all seed hit crystals except those on a calorimeter edge. The spectrum is fit with a Crystal Ball Function, see~\cite{EnerCalNote} for details.}
	\label{Ereso_cos}
\end{figure}

 The energy correction functions $f$ defined above were then applied to the measured cluster energies. The corrected energy of all elastically-scattered electron clusters is shown in Fig.~\ref{Ereso_cos} and used to evaluate the ECal energy resolution as discussed in Sections~\ref{sec:Eres} and~\ref{sec:edges}.

\subsubsection{Calibration with wide-angle Bremsstrahlung events}
The primary physics trigger for two-cluster $e^+e^-$ events also recorded a large number of wide-angle Bremsstrahlung (WAB) events composed of a photon and an electron. These events are selected from two-clusters events, keeping only those with a single matching electron track in the SVT. The sum of energies of these two particles is expected to equal the beam energy. After the calibrations described above, this energy sum was found to be slightly lower, demonstrating that an adjustment of the correction function was needed in the mid-energy range.

 For each WAB event, the energy sum of the two corrected clusters was calculated as :
\begin{equation}
E_{sum} \equiv \frac{E_{e^-}}{f_{e^-}}+\frac{E_{\gamma}}{f_{\gamma}}
\label{eqE}
\end{equation}
where $ E_{i}$ and $f_{i}$  are respectively the cluster energy and the shower leakage correction for the electron or the photon.
Using Eq.~\ref{eqE}, the correction functions $f_{i}$ were adjusted for each particle such that the sum of the two corrected clusters matches the incident beam energy. It was also required that the ratio $f_{e^-}/f_{\gamma}$ be unchanged with respect to the simulation and that the elastically-scattered electrons were not affected. These changes to the energy correction functions were found to be within 1\%.

WAB events provided a reliable method for extracting the energy resolution of the calorimeter at various energies less than the elastic beam energy. A mid-energy point at approximately 0.5~GeV was obtained by selecting WAB events where the energy difference between the two particles was less than 100~MeV.
When selecting only events where both particles are in the fiducial region, the energy resolution can be assumed to be the same for both particles and is obtained by dividing the standard deviation of the energy sum peak by $\sqrt{2}$. A similar procedure was used to find the resolution for highly energy-asymmetric clusters, selecting 0.7 GeV photons in the fiducial region to study the resolution for 0.35 GeV electrons. By understanding the energy resolution of the calorimeter at mid-range energies using WAB events and at the beam energy using elastically-scattered electrons, the full detector response is characterized.

\subsubsection{Gain stability}
\begin{figure}[hbt]
  \centering
      \includegraphics[width=0.8\textwidth]{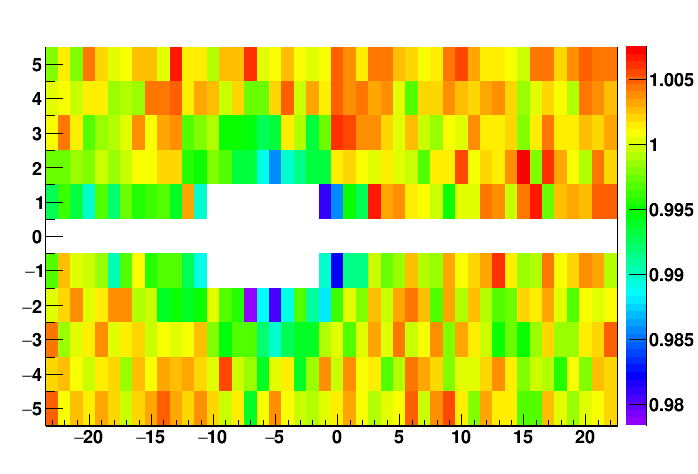}
  \caption{Signal ratio from the LED (LED amplitude after
data taking divided by the initial amplitude), as given by the color scale on the right, after 70 hours of data taking. The axes tick labels refer to the crystals' numbering scheme, ECal being viewed from downstream. }
  \label{gain_stab}
\end{figure}

The gain stability was regularly checked during the runs by comparing amplitudes obtained with LED pulses over time. The response of each channel to LED pulses with a pre-determined amplitude was measured and compared to a reference value.
In preparation, LED settings were equalized to produce a uniform output signal with an amplitude much larger than the channel noise.
The LED signals were checked to be stable to better than 1\% over periods of 3 days without beam.
Figure~\ref{gain_stab} shows the ratio of LED signal after and before 70 hours of data taking with beam on target at the nominal luminosity.
Most of the crystal gains were stable to within 1\%.
Around the vacuum vessel, where the particle rate was much higher, the gains are slightly reduced.

\subsubsection{Energy resolution in the fiducial region}
\label{sec:Eres}
\begin{figure}[hbt]
  \centering
      \includegraphics[width=0.7\textwidth]{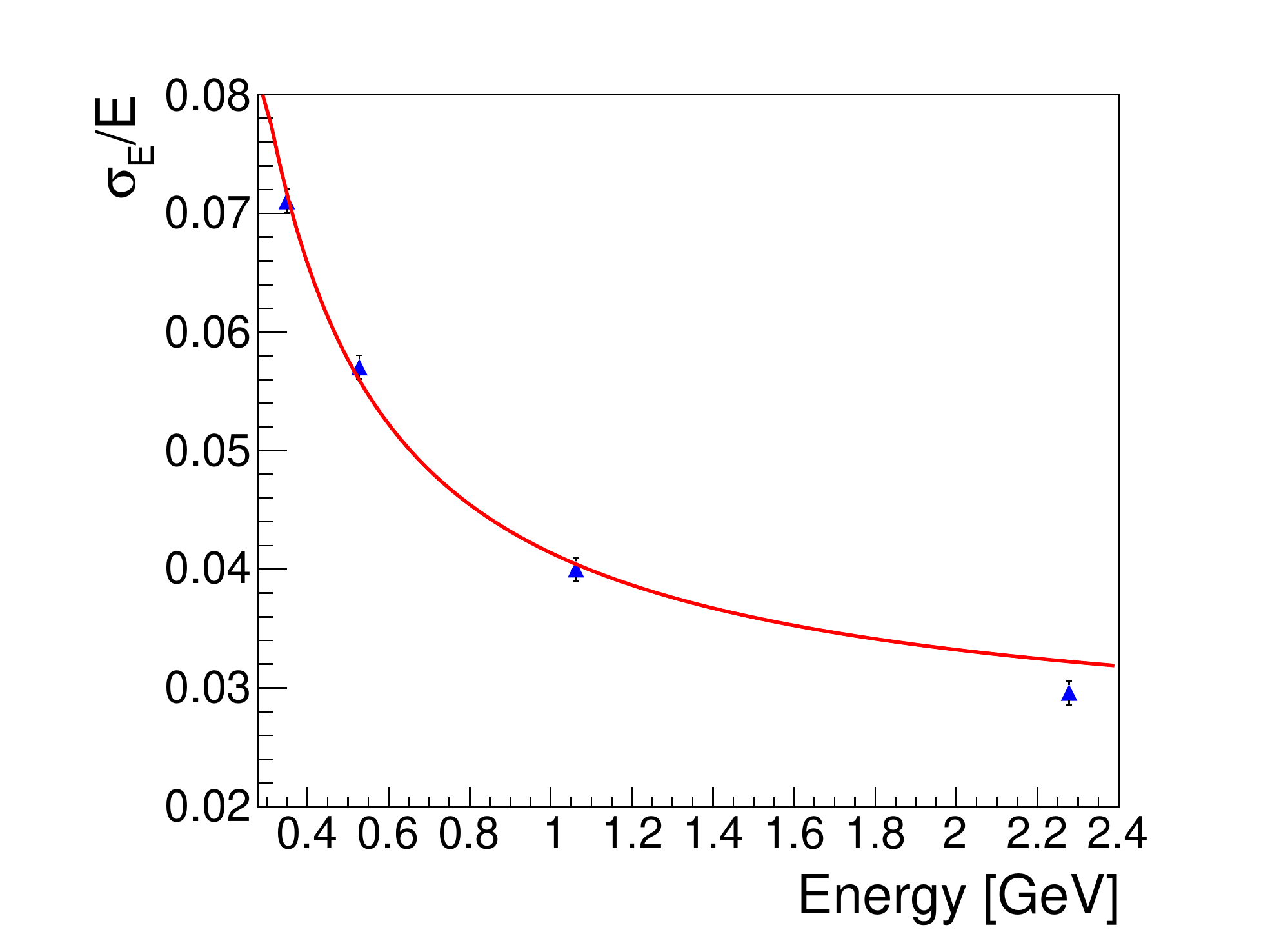}
  \caption{Energy resolution in the ECal as found in data by using elastically-scattered electrons and WAB events. The fit is given by Eq.~\ref{eq:enres} and only used data at 1.05~GeV beam energy. The point at 2.3~GeV was not included in the resolution fit; it was obtained using 2.3~GeV beam energy. }
  \label{eres}
\end{figure}

 The calibration with elastically-scattered electrons provided the cleanest point in understanding the energy resolution at the beam energy.
WAB events were used to assess the energy resolution at mid-range energies. By cutting on  the energy of the photons, the energy resolution of electrons was studied both as a function of energy and position relative to the edges of the ECal. The resulting energy resolution in the central region of the calorimeter is shown in Fig.~\ref{eres}.

The fit to the energy resolution results in:
\begin{equation}
\label{eq:enres}
\frac{\sigma_E}{E}(\%)=\frac{1.62}{E}\oplus \frac{2.87}{\sqrt{E}} \oplus 2.5,
\end{equation}
 where the $\oplus $ symbol indicates a quadratic sum and $E$ is in units of GeV.
 In Eq.~\ref{eq:enres}, the first term is generally attributed to electronic noise. The second term is related to the statistical fluctuations of the shower containment and the APD gain. The last term contains both the energy leakage out the back of the ECal and the crystal-to-crystal inter-calibration error.


\subsection{Edge effects}
\label{sec:edges}
\begin{figure}[thb]
  \centering
      \includegraphics[width=1.0\textwidth]{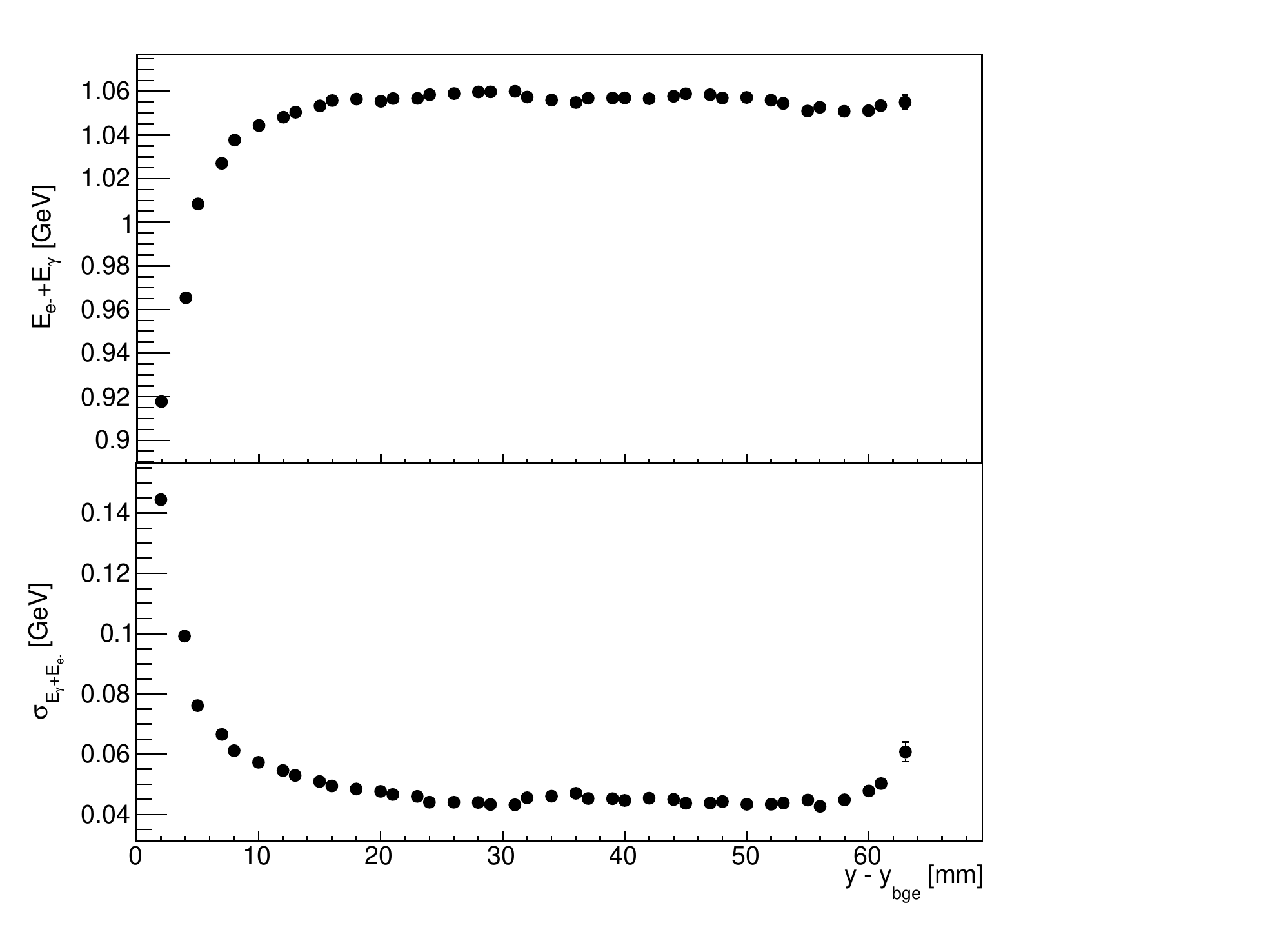}
  \caption{WAB events with $|E_{e^-}-E_{\gamma} | < 0.1$ GeV.
  Top: energy sum ($E_{e^-}+E_{\gamma}$) peak position.
  Bottom: energy sum peak standard deviation $\sigma_{E_{e^-}+E_\gamma}(y)$ as a function of vertical position of the electron across the ECal.}
  \label{wabHalf}
\end{figure}

To understand the energy resolution of clusters at the edges of the calorimeter, a study was performed using
events from both WAB and elastically-scattered electrons.
The electron position at the ECal, given by the SVT track, was used to determine the electron's distance from the beam gap edge.
In the case of WAB events, the photon was required to be within the ECal fiducial region.
The energy resolution dependence on the electron vertical position could be obtained from the data by relaxing the condition on the electron to be in the fiducial region (see Fig.~\ref{wabHalf}):
 \begin{equation}
\label{eq:WAB_edges}
\sigma_{{\scriptscriptstyle E_{e^-}}}(y)=\sqrt{\sigma^2_{{\scriptscriptstyle E_{e^-}+E_{\gamma}}}(y)-\sigma^2_{{\scriptscriptstyle E\gamma}}(y_0)} ,
\end{equation}
where $y_0$ is in the fiducial region. The photon is required to be in the fiducial region and its energy is therefore well-measured. Selecting the energy of the photon determines the electron's energy since their sum equals the beam energy. In this way, the energy resolution of the electron was determined as a function of its distance from the beam gap edge of the calorimeter.

\begin{figure}[thb]
  \centering
      \includegraphics[width=0.7\textwidth]{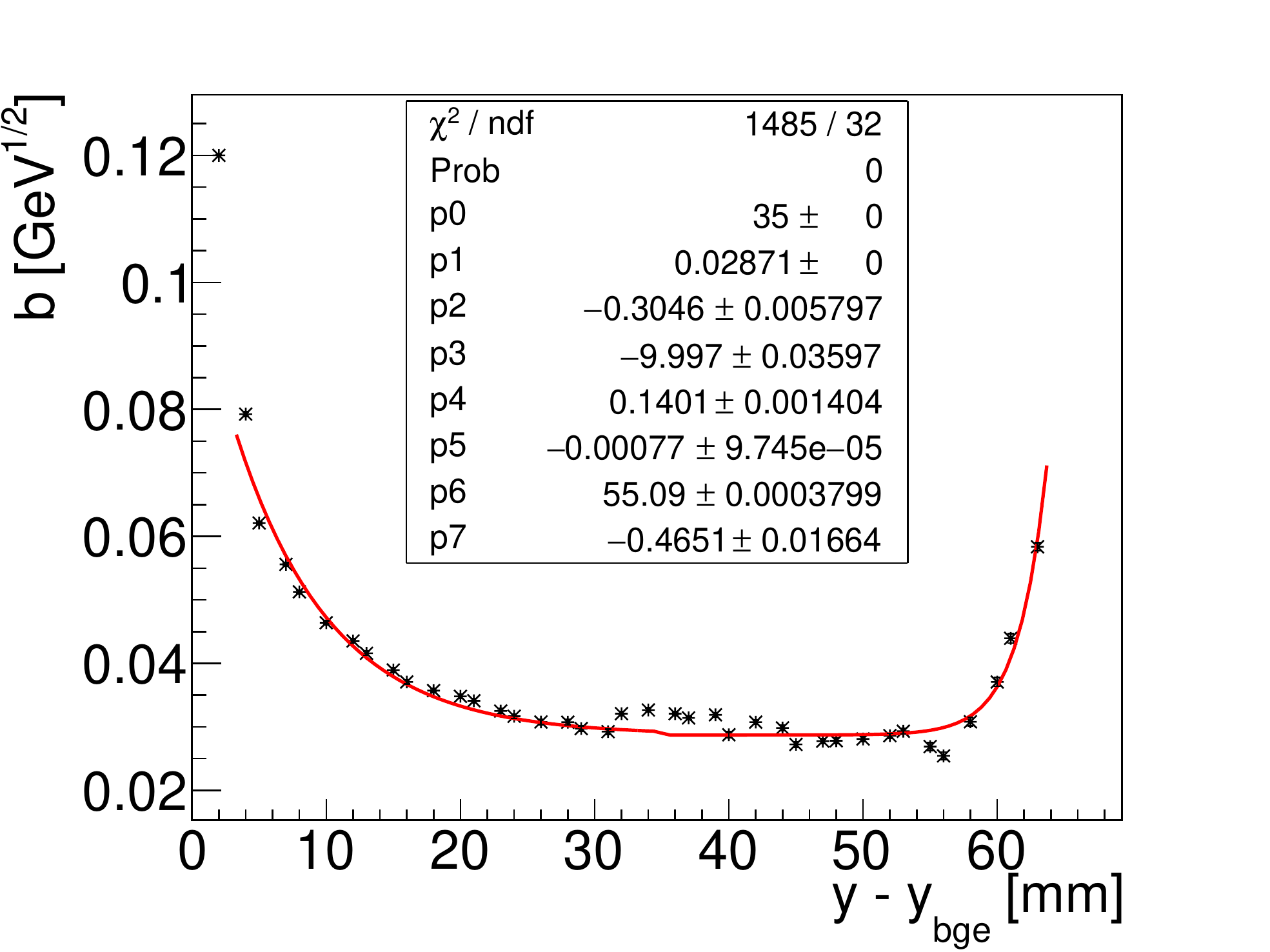}
  \caption{The stochastic parameter $b$ (corresponding to the $1/\sqrt{E}$ term) of the energy resolution description is shown as a function of the vertical position relative to the ECal beam gap edge.
                 The fit function is given in Eq.~\ref{eq:finalRes}.
                 }
  \label{BparRes}
\end{figure}

It was found that the second parameter of the energy resolution function, $b/\sqrt{E}$, is strongly correlated to the particle position in the ECal. By fixing the other two parameters to the values as found in the fiducial region (see Eq.~\ref{eq:enres}), the $b$ parameter was determined as a function of position, as illustrated in Fig.~\ref{BparRes}.
 The ECal energy resolution dependence on both the energy and vertical position is then simply given by:
 \begin{equation}
\label{eq:finalRes}
\frac{\sigma_E}{E}(\%)=\frac{1.62}{E}\oplus \frac{b(y-y_{bge})}{\sqrt{E}} \oplus 2.5
\end{equation}
where, $y_{bge}$ is the $y$ position at the inner gap edge. Similarly to previous such fits at the edges, two matching exponentials were used to parameterize $b$: \\
$b(|y-y_{bge}|<p_0) = p_1 - p_2 e^{-(y-p_3) \cdot p_4}$, \\
$b(|y-y_{bge}|>p_0) = p_1 - p_5 e^{-(y-p_6) \cdot p_7}$. \\
The energy resolution deteriorates rapidly within 8-10~mm from the edge of the calorimeter. Equation~\ref{eq:finalRes} is however reliable down to 6.5~mm (which corresponds to the center of the last crystal front face) from the edge of the calorimeter.


\subsection{Time calibration}
HPS is a high rate experiment, up to 1~MHz per crystal and 30~MHz for the whole calorimeter,
with a 15~MeV threshold per crystal during typical run conditions.
The time calibration is a key element for reducing backgrounds for accidentals and determining which clusters are coincident.

\subsubsection{Crystal pulse fitting}
\label{sec:PulseFitting}
\begin{figure}[ht!]
\centering
\includegraphics[width=0.65\textwidth]{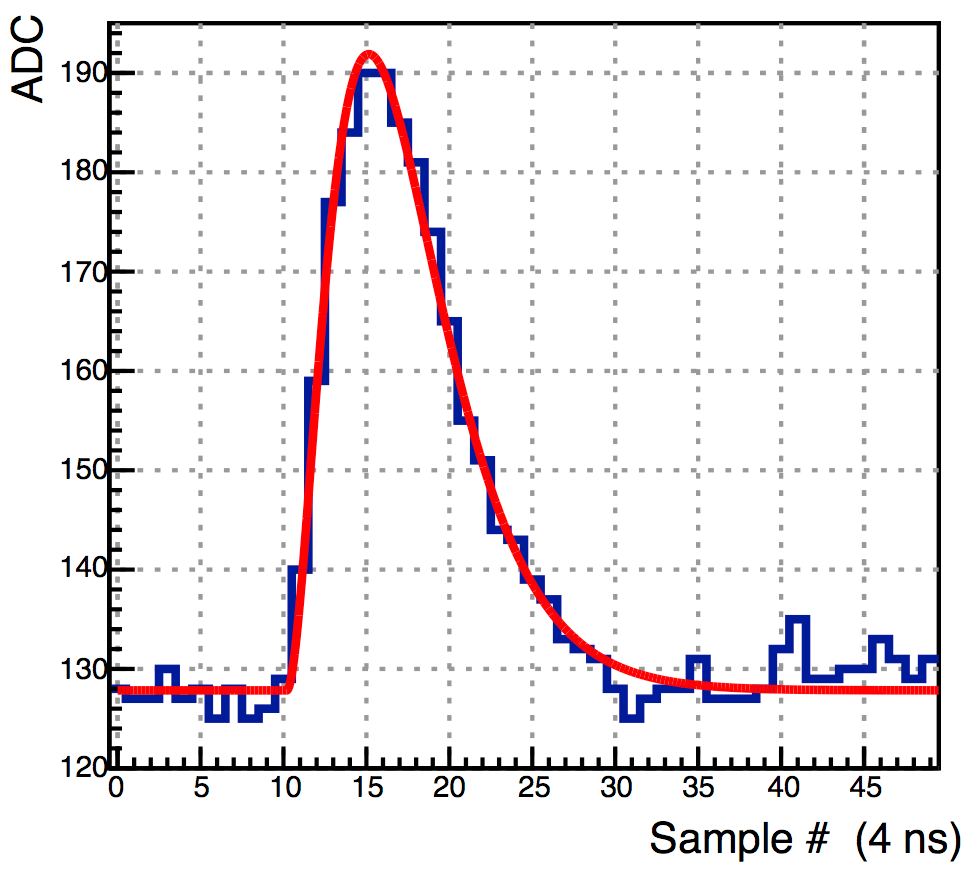}
\caption{Example fit of an individual crystal pulse.
		}
\label{PulseFit}
\end{figure}

The time of a cluster is set from the seed crystal (the crystal with the highest energy  in the cluster).
This crystal pulse shape, sampled at 250~MHz by the FADC, is
fit by the sum of a pedestal $P$ and a 3-pole function with
width $\tau$ and time $t_{0}$~\cite{Andrea-Gabriel-Note}:
\begin{equation}
\textrm{ADC}(t) = P + \frac{A}{2 \tau^{2}} \left( t-t_{0} \right)^{2} e^{- \left( t-t_{0} \right) / \tau}
\label{3pole}
\end{equation}
An example fit is shown in Fig.~\ref{PulseFit}.
Best resolutions were obtained by fixing, for each crystal independently, the width parameter to the average value measured over many pulses~\cite{NathansNote}.


\subsubsection{Crystal time alignment}
Corrections to the above defined pulse time $t_0$ are needed in order to remove timing variations from crystal to crystal:
\begin{equation}
t = t_0 + \Delta t_{RF} + \Delta t_{tw}(E),
\label{eqToff}
\end{equation}
where $\Delta t_{tw}(E)$ is the time walk correction and $\Delta t_{RF}$ is the channel time offset with respect to the RF signal obtained before time walk corrections. The two quantities are  discussed in the following.

\begin{figure}[ht!]
\centering
\includegraphics[width=0.65\textwidth]{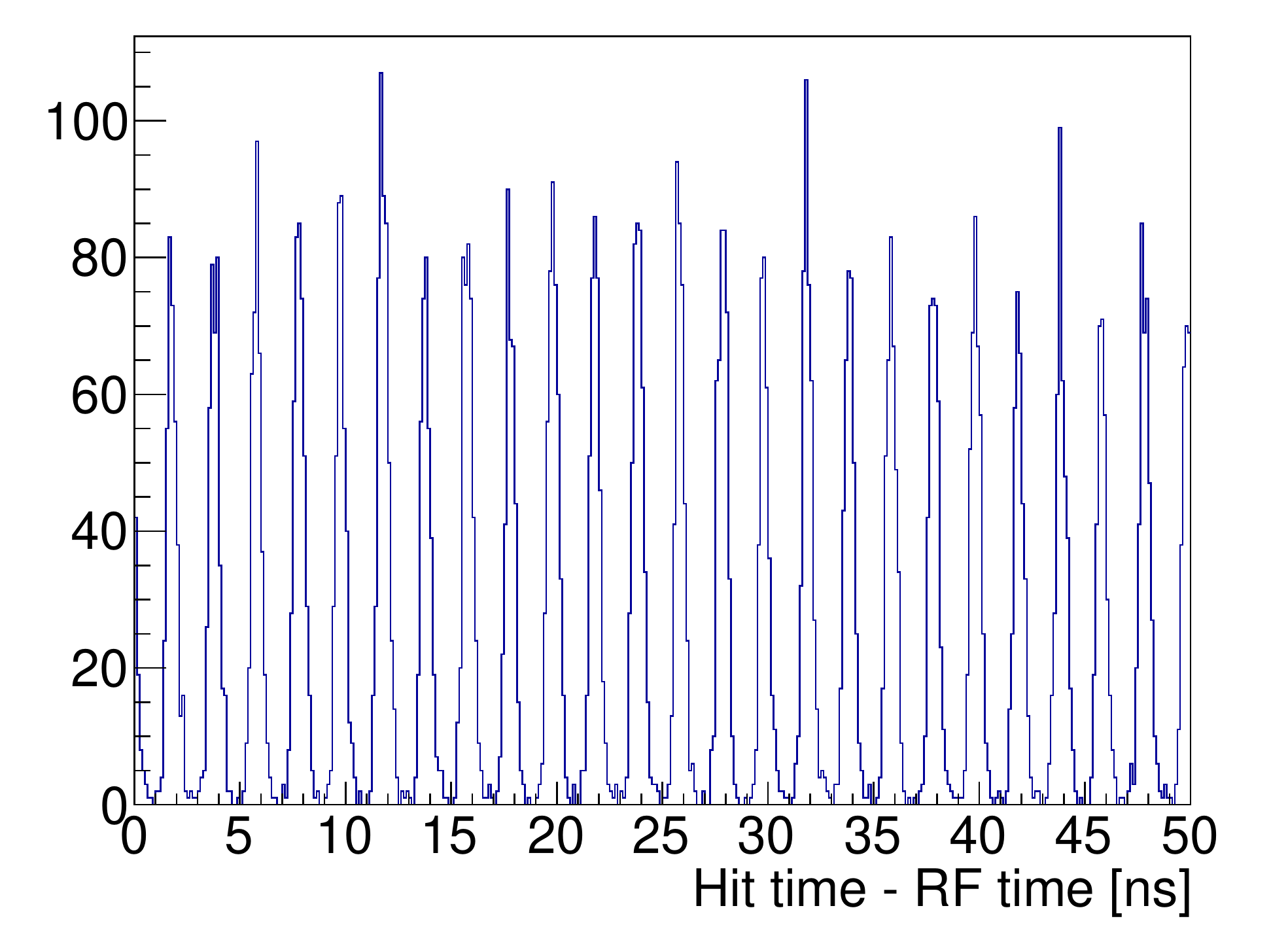}
\caption{Time difference between a single crystal hit and the RF time.}
\label{RF}
\end{figure}

The accelerator 499~MHz RF signal, measured with a FADC channel in the same conditions as the ECal signals, is sampled in one of every 80 accelerator beam bunches.
The precision at which this signal is measured has been determined to be 24~ps.
Over many events, the accelerator beam structure is clearly observed in the distribution of the time difference between a hit crystal in a cluster and the accelerator RF signal time, resulting in peaks evenly spaced every 2.004~ns as seen in Fig.~\ref{RF}.

\begin{figure}[ht!]
\centering
\includegraphics[width=0.48\textwidth]{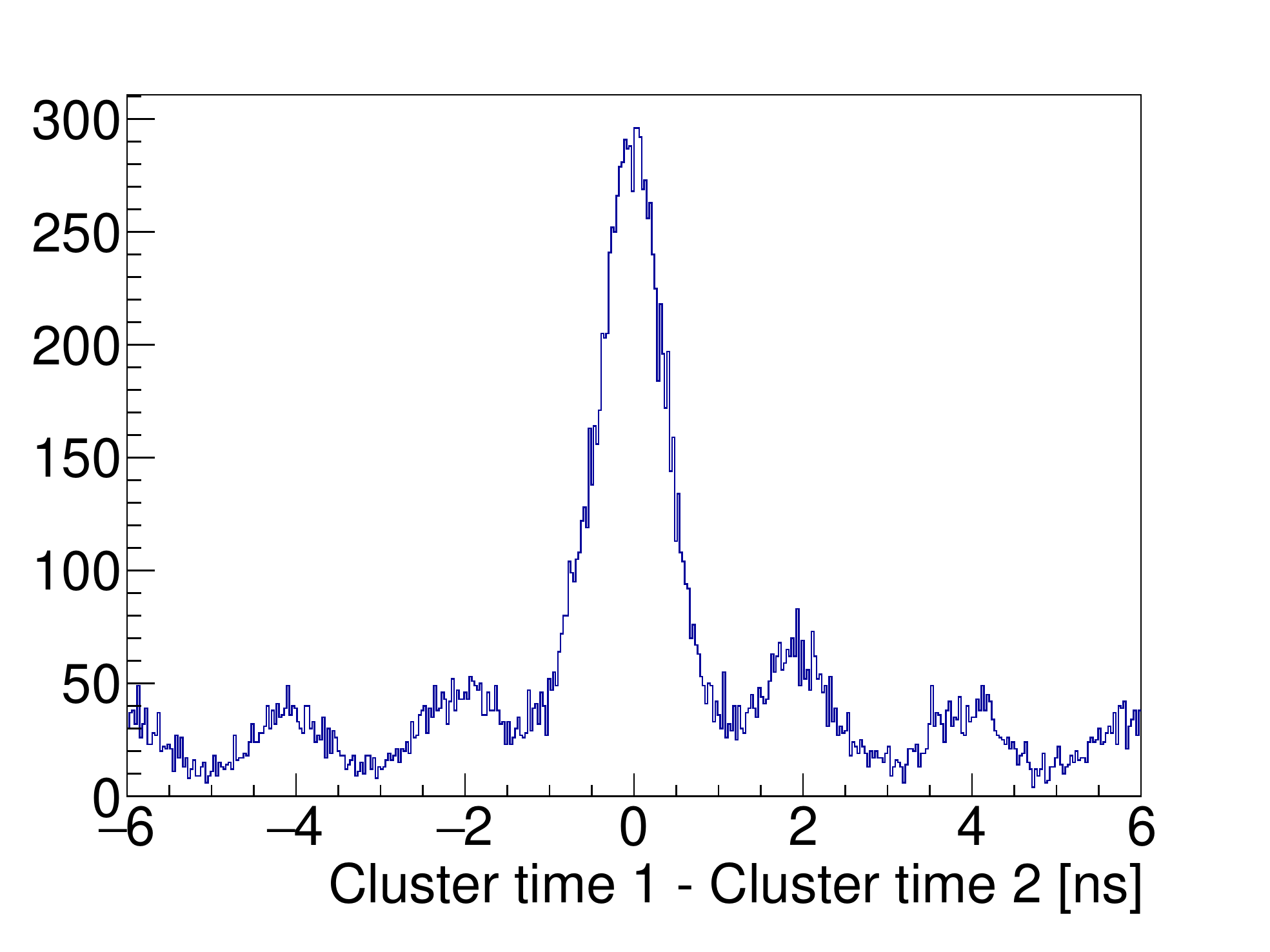}
\includegraphics[width=0.49\textwidth]{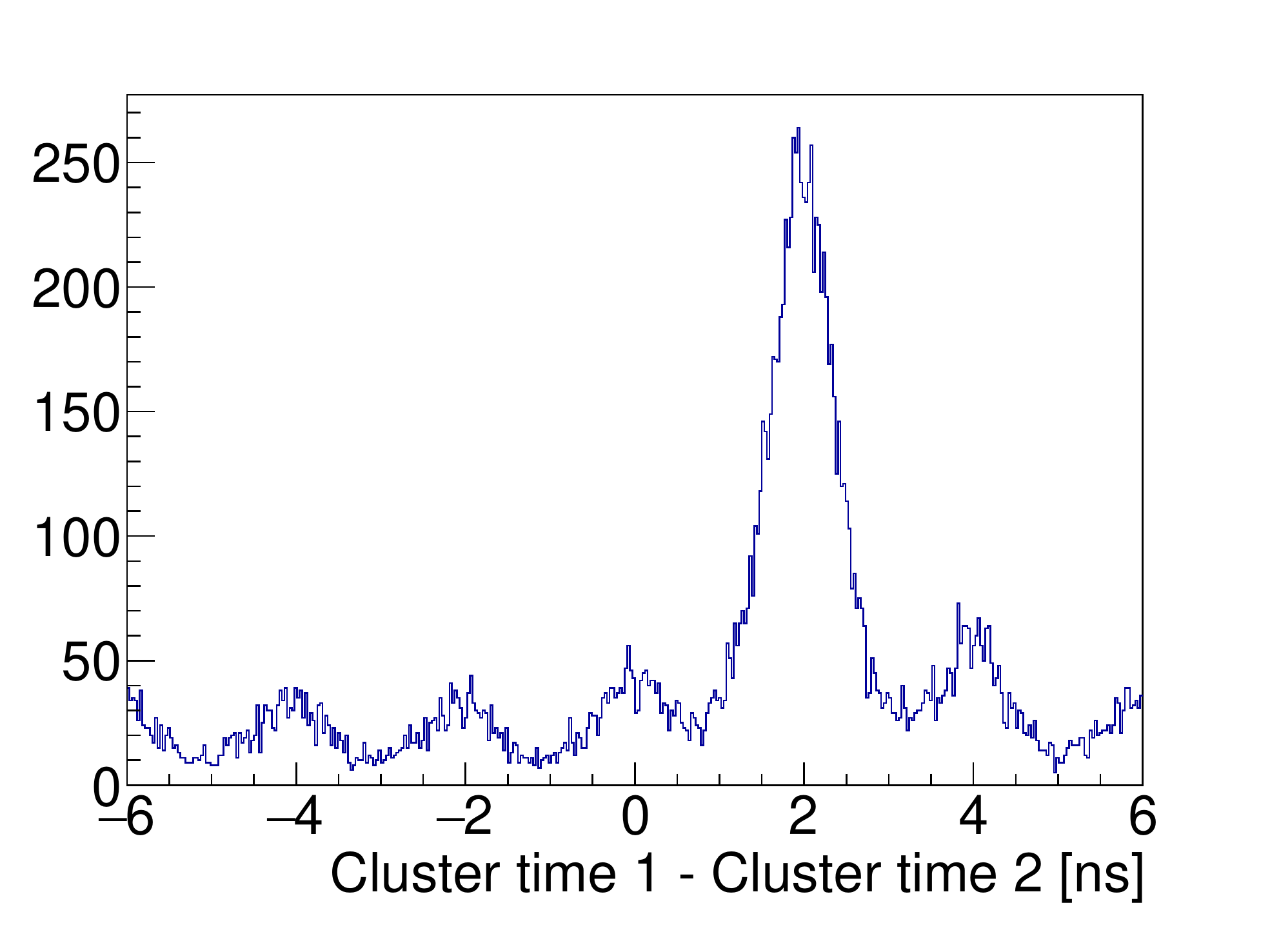}
\caption{\label{LargeOffset}
		Time difference between the seed hits of two correlated clusters after finding initial RF offsets.
		Cluster 1 has a given (fixed over all events) seed crystal while cluster 2 may be anywhere in the ECal.
		The left and right plots show events for two different choices of seed crystals for cluster 1.
		}
\end{figure}

Using the time difference between the hit time with the RF signal time, the time offsets of each channel could be fine-tuned for scales smaller than the 2.004~ns spacing:
\begin{equation}
\Delta t_{fine} = modulo(t_{hit} - t_{RF} +N\times2.004, 2.004),
\label{fineTime}
\end{equation}
where $N$ is an arbitrarily large integer that ensures the time difference is positive. Once the $\Delta t_{fine}$ offsets have been found, the crystals have large time offsets in increments of 2.004~ns (resulting from the use of the modulo with the $t_{RF}$) that must be determined. These large offsets were found by studying well-correlated two cluster events and measuring the time difference between the seed hits of the two clusters as shown in Fig.~\ref{LargeOffset}.

For these events, correlated clusters are required to have an energy sum close to the beam energy,
an energy difference less than 200~MeV, and occur within a given trigger time window.
The time of the largest peak indicates the offset in increments of 2.004~ns that should be applied in addition to the previous $\Delta t_{fine}$ offset in order to obtain $\Delta t_{RF}$.

\begin{figure}[bht]
\centering
\includegraphics[width=0.7\textwidth]{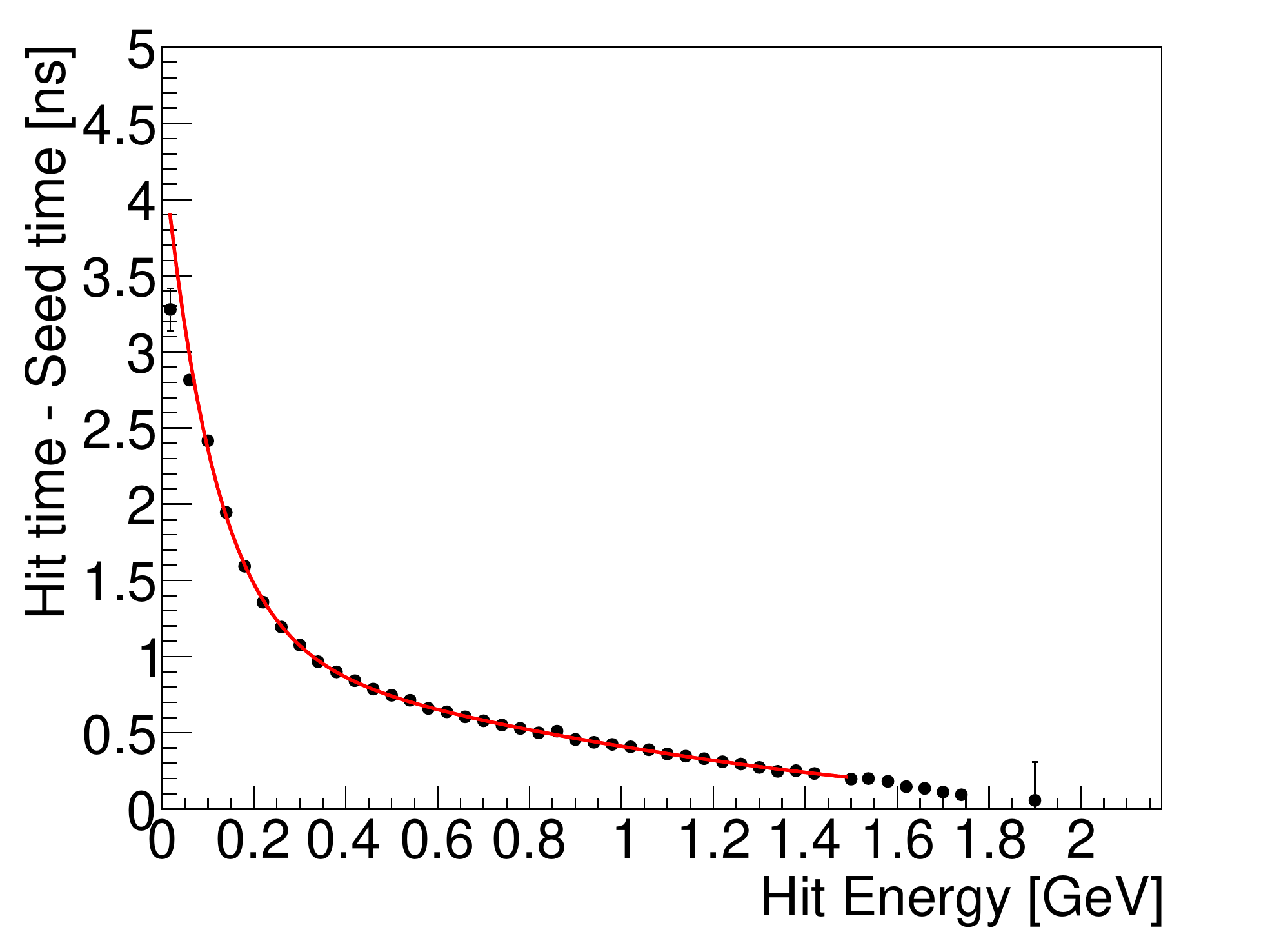}
\caption{Time walk correction as a function of hit energy (within clusters where the seed hit energy is greater than 400~MeV).}
\label{TimeWalk}
\end{figure}

After aligning all crystal offsets in time, the energy dependence of the time offsets, or time walk, can be characterized. The time walk is measured using the time differences
between individual hits in a cluster and the highest energy hit as a function of hit energy. The
results are fitted by an exponential and a second order polynomial,
as shown in Fig.~\ref{TimeWalk}, and form the basis of a time-walk correction.
The time walk is very small for crystal energies above 150~MeV, and thus does not significantly
affect the resolution of the two-cluster time difference. It is however important
for the time offsets in the clustering algorithm in order to enable tighter time cuts between crystals.

\subsubsection{Time resolution}
\begin{figure}[ht!]
\centering
\includegraphics[width=0.7\textwidth]{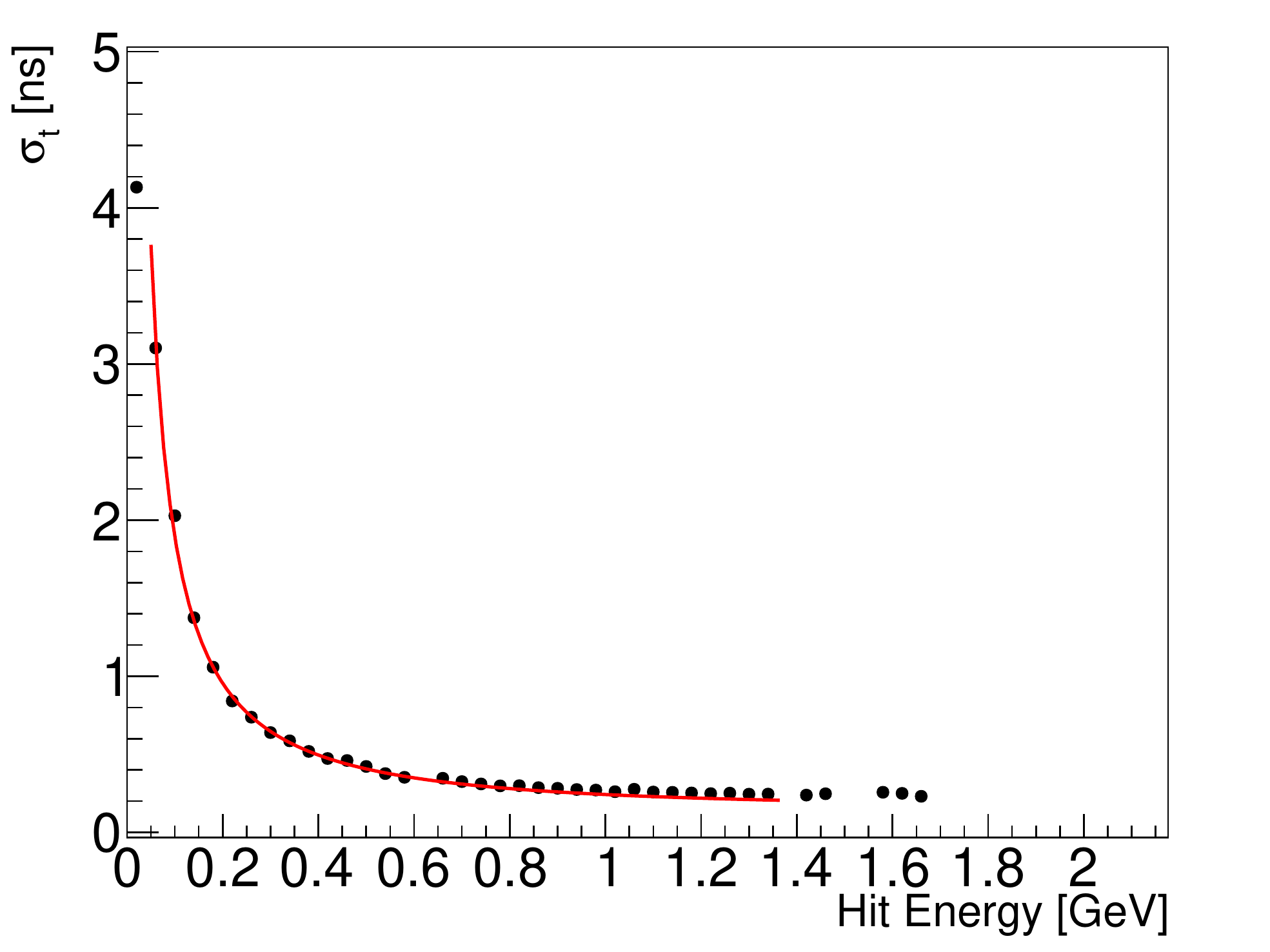}
\caption{Individual crystal time resolution as a function of energy.}
\label{TresoVSTime}
\end{figure}

Finally, the time resolution as a function of hit energy is extracted from
the width of the time coincidences within single clusters. The result is shown in Fig.~\ref{TresoVSTime}, and
the time resolution can be parameterized as~\cite{TimeCalNote}:
\begin{equation}
\textrm{Time resolution}~\textrm{(ns)}= \frac{0.188}{E~(\textrm{GeV})} \oplus 0.152.
\label{eq}
\end{equation}

The obtained time resolution is significantly smaller than the intrinsic 4~ns FADC sampling period and enables the use of the ECal to improve offline event selection and reduce accidentals from the final analysis.

\section{Trigger performance}
\label{sec:Trigger}
The electromagnetic calorimeter is the only detector used in the HPS trigger decision.
Therefore a significant effort has been made to ensure that it has an efficient background rejection while maximizing the acceptance for $A^\prime$ events.

\subsection{General scheme of the trigger}
The HPS trigger scheme is as follows. The analog signal from each ECal channel is continuously sampled by the FADC every 4~ns. When the signal crosses a selectable threshold, N$_1$ samples before crossing and N$_2$ samples after (typically 5 and 25) are summed together to provide the pulse charge, which
is then converted into energy, using online gains and pedestals loaded, channel by channel, in the FADC. The resulting energy and threshold crossing time are then passed every 16~ns to the clustering algorithm in the General Trigger Processor board (GTP).

The first step of processing in the GTP is finding seed crystals.  A hit
is considered to be a seed if it fulfills two conditions: an energy higher than a selectable threshold and higher than all its 8 nearest neighbors (or fewer neighbors if it belongs to one of the calorimeter edges).

When one of the crystal energies meets the definition of a seed hit, a time coincidence between the seed hit and its neighbors is then required to group additional hits into the cluster.
The timing coincidence is programmable, typically 4 samples, and required to compensate for time-walk effects.
The cluster energy is the sum of all the crystal energies within a 3$\times$3 spatial array and time constraints.
Once the clustering algorithm on the GTP has identified a cluster, the corresponding data is reported to the main trigger processor.
This includes: the timestamp, the energy, and the spatial coordinates (center of the seed crystal).
The cluster energy is not corrected for shower leakage effects at this stage.
Finally, the trigger processor makes the trigger decision by applying further selection to the clusters.
Currently, two event topologies are considered, with one or two clusters.
Parallel trigger selections are implemented in the trigger processor, with an associated prescale factor of $2^{n-1}+1$, where $n$ is selectable.

\subsection{Trigger parameters}
The system includes two pair triggers, two single cluster triggers, a random pulser trigger for background studies and a calibration trigger for cosmics and the LED monitoring system. All these triggers have separate trigger cuts and can operate simultaneously with individual prescale factors.

The single cluster triggers are based on lower and upper energy limits and a number of hits in the cluster.
One of the single cluster triggers was tuned to select the elastic scattering of electrons off the nuclear target. The second did not have stringent cuts and serves for testing purposes, in particular for the determination of trigger efficiencies.

The two cluster pairs triggers were optimized for different physical processes and used different sets of parameters. The Pair-0 trigger algorithm was used for the selection of M\o ller scattering, while Pair-1 was the main trigger for the Heavy Photon search. Each trigger, except Pair-1, has an associated prescale factor, in order to keep the total trigger rate acceptable for the data acquisition system.

The cuts applied for the main trigger, cluster pairs, are presented below.
Cluster pairs are generated by forming all possible combinations of clusters from the top and the bottom half of the calorimeter. There are seven cluster-pair cuts.
Denoting cluster energy, number of hits, time, and coordinates as $E_{i}, N_{i}, t_{i}, x_{i}, y_{i}$, where $i=1$~or~$2$, the cuts are defined as:
\begin{equation}
E_{\textrm{min}} \leq E_i  \leq E_{\textrm{max}},
\end{equation}
\begin{equation}
E_{\textrm{sum min}} \leq E_{1} + E_{2} \leq E_{\textrm{sum max}},
\end{equation}
\begin{equation}
 N_{i} \geq N_{\textrm{threshold}},
\end{equation}
\begin{equation}
E_2 - E_1  \leq E_{\textrm{difference}},
\end{equation}
\begin{equation}
E_1 + r_1 F \geq E_{\textrm{slope}},
\end{equation}
\begin{equation}
|\arctan{\frac{x_{1}}{y_{1}}} - \arctan{\frac{x_{2}}{y_{2}}} | \leq \theta_{\textrm{coplanarity}},
\end{equation}
\begin{equation}
|t_{1} - t_{2}| \leq t_{\textrm{coincidence}}.
\end{equation}
\noindent
$E_{\textrm{min}}$, $E_{\textrm{max}}$, $E_{\textrm{sum min}}$, $E_{\textrm{sum max}}$, $E_{\textrm{difference}}$, $E_{\textrm{slope}}$, $F$, $\theta_{\textrm{coplanarity}}$, $N_{\textrm{threshold}}$ and $t_{\textrm{coincidence}}$ are programmable trigger parameters.
$E_1$ is the energy of the cluster with the lowest energy and
 $r_1=\sqrt{x_1^2+y_1^2}$ is the distance between its center and the calorimeter center.

\begin{figure}[ht!]
\centering
\includegraphics[width=0.95\textwidth]{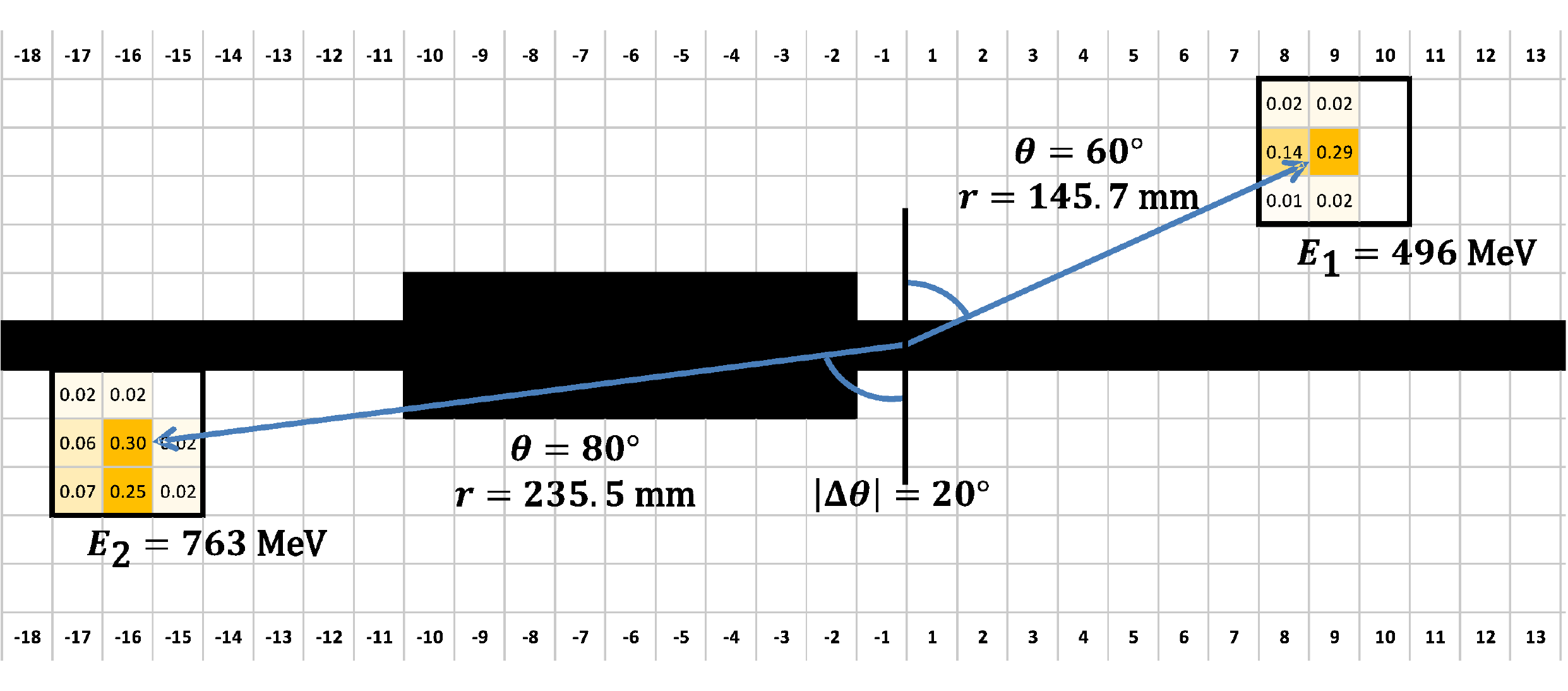}
\caption{Real event selected by the HPS trigger.
		The event includes two clusters with energies
		$E_1=496$~MeV at the top part of the calorimeter and $E_2=763$~MeV at the bottom half.
		The corresponding $3\times3$ spatial windows are shown.
		The distances $r_i$ between seed crystals and the center of the calorimeter are indicated.
		The coplanarity angle is calculated as $|\theta_1-\theta_2|=80^\circ -60^\circ=20^{\circ}$.}
\label{Trigger}
\end{figure}

The values chosen for each parameter were based on Monte Carlo simulations and are summarized in Table~\ref{tab:pair_trigger_settings}.
The Pair-1 trigger parameters were tuned for each beam energy and chosen to ensure the best $A^\prime$ signal efficiency
and signal-over-background ratio. Note that most Pair-1 events have at least one cluster on a calorimeter edge. The degraded energy response at the edges is taken into account in the simulations that led to the choice of selection cuts. Figure~\ref{Trigger} shows a real event selected by the HPS pair trigger algorithm.
\begin{table}[h!]
	\begin{center}
		\begin{tabular}{lcccc}
			\hline
			\textbf{Parameter} & \textbf{Single-0} & \textbf{Single-1} & \textbf{Pair-0} & \textbf{Pair-1}\\
			\hline
			\( E_{\textrm{min}} \)                   & 0.100 GeV   & 1.300 GeV & 0.150 GeV     & 0.150 GeV\\
			\( E_{\textrm{max}} \)                 & 2.700 GeV   & 2.600 GeV & 1.400 GeV      & 1.400 GeV\\
			\( N_{\textrm{threshold}} \)         & 3 hits            & 3 hits          & 2 hits              & 2 hits\\
			\( E_{\textrm{sum min}} \)           & ---                 & ---               & 0.500 GeV     & 0.600 GeV\\
			\( E_{\textrm{sum max}} \)          & ---                 & ---               & 1.900 GeV     & 2.000 GeV\\
			\( E_{\textrm{difference}} \)         &---                  & ---               & 1.100 GeV     & 1.100 GeV\\
			\( E_{\textrm{slope}} \)                & ---                 & ---                & 0.400 GeV     & 0.600 GeV\\
			\( F \)                                           &---                  &---                 & 0.0055 GeV/mm   & 0.0055 GeV/mm\\
			\( \theta_{\textrm{coplanarity}} \) & ---                 & ---                & ---                   & 40\textdegree \\
			\( t_{\textrm{coincidence}} \)       &  ---                &      ---           & 8 ns                & 12 ns\\
			Prescale                                     & $2^{12}+1$       &  $2^{10}+1$     & $2^{5}+1 $        & $1$ \\
			\hline
		\end{tabular}
		\caption{All trigger settings for the Single-0, Single-1, Pair-0 and Pair-1 triggers for the run with beam energy 2.3~GeV.
				Note that energies are not corrected for shower leakage at the trigger stage. The purpose of the different triggers is described in the text.}
		\label{tab:pair_trigger_settings}
	\end{center}
\end{table}

\subsection{Trigger diagnostics}
The main diagnostic consists of a comparison between the hardware trigger algorithm and its software simulation. The numbers of clusters and triggers are compared with both the hardware and the simulated triggers. The results show an agreement above 99\%.
The small difference is due to fluctuations for near-threshold energy in the selection of the clusters and for events at the edge of the time window.



\begin{figure}[htbp]\centering
    \includegraphics[width=0.60\textwidth]{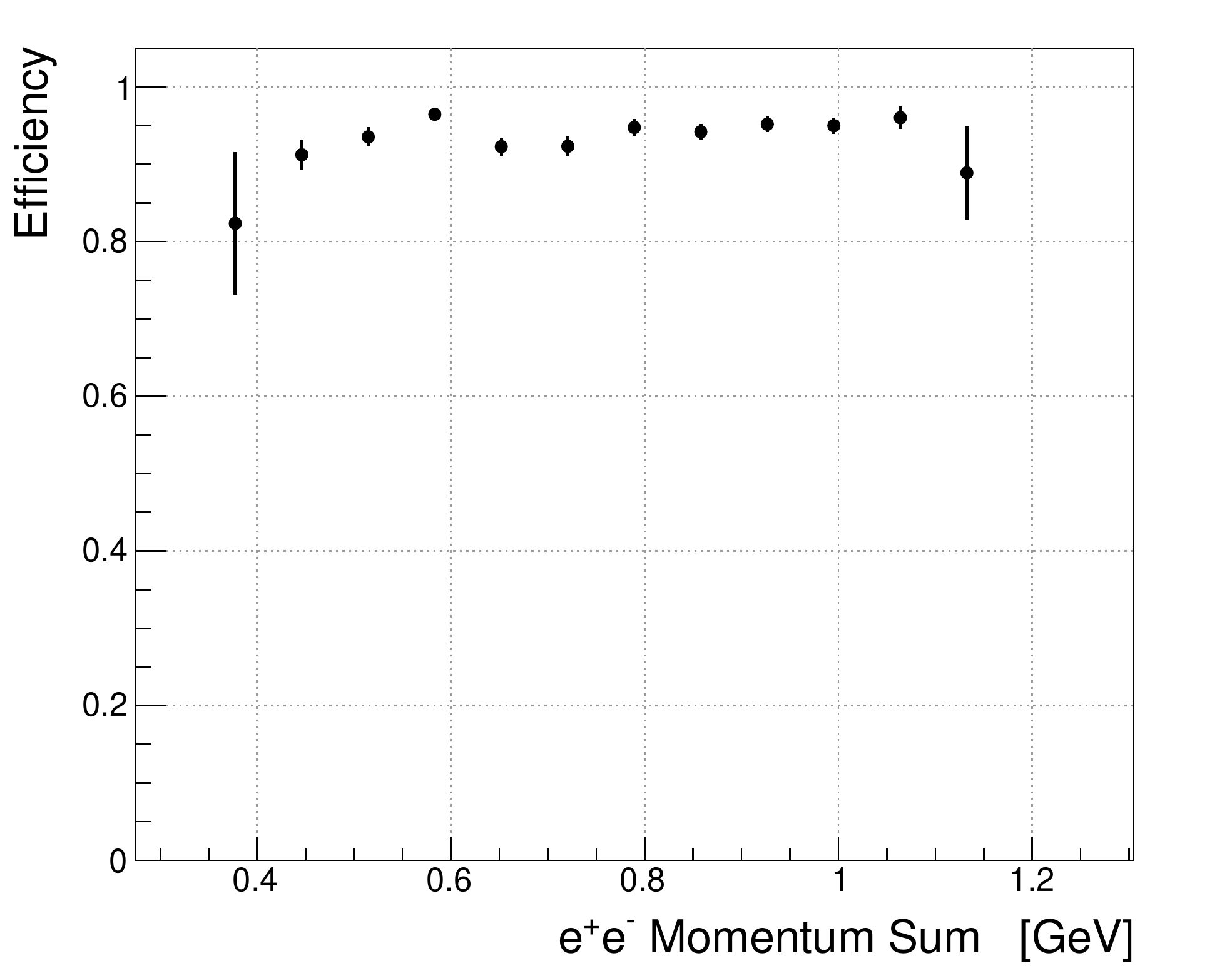}
    \caption{\label{fig:tritrigeff}An effective $e^+e^-$ Pair-1 trigger efficiency measured from the 2015 run with 1.056~GeV beam energy based on random pulser triggers and tracks projected to the calorimeter.}
\end{figure}

An effective $e^+e^-$ Pair-1 trigger efficiency based on comparing reconstructed tracks and recorded trigger bits is also measured.  Events are first selected from random pulser triggers with at least two oppositely charged tracks in the SVT, one in each half of the detector.  The tracks must be well measured 
with small $\chi^2$ and project to the calorimeter fiducial region (at least 1.5 mm in from its edge).  Track pairs are then required to satisfy Pair-1 conditions corresponding to those in Table~\ref{tab:pair_trigger_settings}, but with energies adjusted to compensate for differences in tracking and calorimeter measurements.  The fraction of those events that also have a Pair-1 trigger bit set from the calorimeter is shown in Figure~\ref{fig:tritrigeff}, illustrating a very high trigger efficiency.

\section{Cluster-Track matching}
\label{sec:SVTmatching}
\subsection{Need for cluster-track matching}
The cluster-track matching is an important part of the background
reduction in the physics analysis. In addition, as described in Section \ref{EnerCalibSubsection}, 
cluster energy and position corrections depend on particle type. Therefore, 
it is necessary to know whether each cluster is associated with $e^-$, $e^+$, or photon
before applying the corrections.
This can be determined based on matching the clusters with tracks, or, in the case of photons,
the lack of an associated track.

Tracks reconstructed with the SVT can be extrapolated to the calorimeter to determine their intersection with
the ECal. The residual between the reconstructed cluster
and the extrapolated track coordinates
is used as a measure of cluster-track matching.


\subsection{Selection of samples and determination/parametrization of matching functions}
To develop the matching criteria, a strict event selection was applied. Two time-coincident clusters in the ECal and two oppositely charged tracks in the SVT, one in each half of the detector, were required.
\begin{figure}[!htb]
 \centering
 \includegraphics[width=0.65\textwidth]{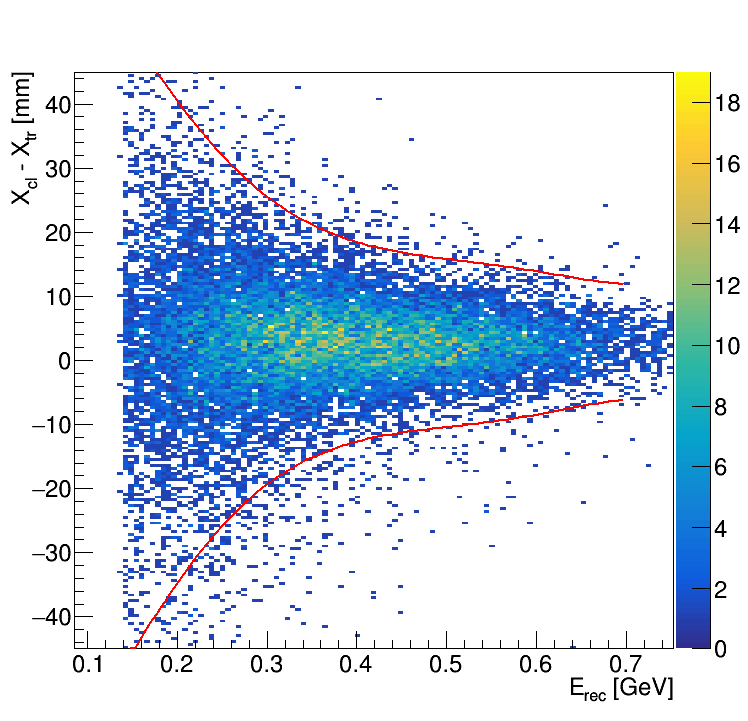}
 \caption{Difference of cluster and track horizontal coordinate ($x$) as a function of uncorrected cluster energy,
 for ECal bottom half and negative tracks. Red contour lines represent our parameterizations of $\mu\pm3\sigma$.}
 \label{fig:dX_Ecl_bot_pos}
\end{figure}

Due to the possibility of small misalignments of the detector and of the dipole magnet, there can be independent systematic shifts for the two detector halves and particle charges. Therefore the cluster-track matching is 
studied separately for all combinations of bottom-top and negative-positive tracks.
Since cluster and extrapolated track position resolutions are energy dependent, it is also natural to parametrize the matching as a function
of energy. 

Figure \ref{fig:dX_Ecl_bot_pos} shows an example of the horizontal coordinate 
difference between clusters and negative tracks as a function of the uncorrected cluster energy in the bottom half of the detector. This figure also illustrates that this selection is essentially free of background.

In total, there are eight similar distributions:
$2$ (coordinate) $\times\;2$ (detector half) $\times\;2$ (track charge).
All eight were divided into 20 slices of energy, and, 
for each slice the residuals were fit with a Gaussian function. The Gaussian means $\mu$ and widths $\sigma$ were then parameterized with a $5^{th}$ degree
polynomial as a function of energy. 
The red contour lines in Figure \ref{fig:dX_Ecl_bot_pos} show an example of these $\mu \pm 3\sigma$ functions. 

To quantify the degree of matching for a particular cluster and track with measured $x$ and $y$ positions, we define a quantity $n_\sigma$ as 
\begin{equation}
    n_\sigma = \sqrt{n_{\sigma_{x}}^{2} + n_{\sigma_{y}}^{2}},
\end{equation}
where
\begin{equation}
n_{\sigma_{x}} =\frac{x_{cluster} - x_{track} - \mu_{x}}{\sigma_{x}},
 \end{equation}
 and similarly for the $y$-coordinate.

\begin{figure}
   \centering
\includegraphics[height=0.50\textwidth]{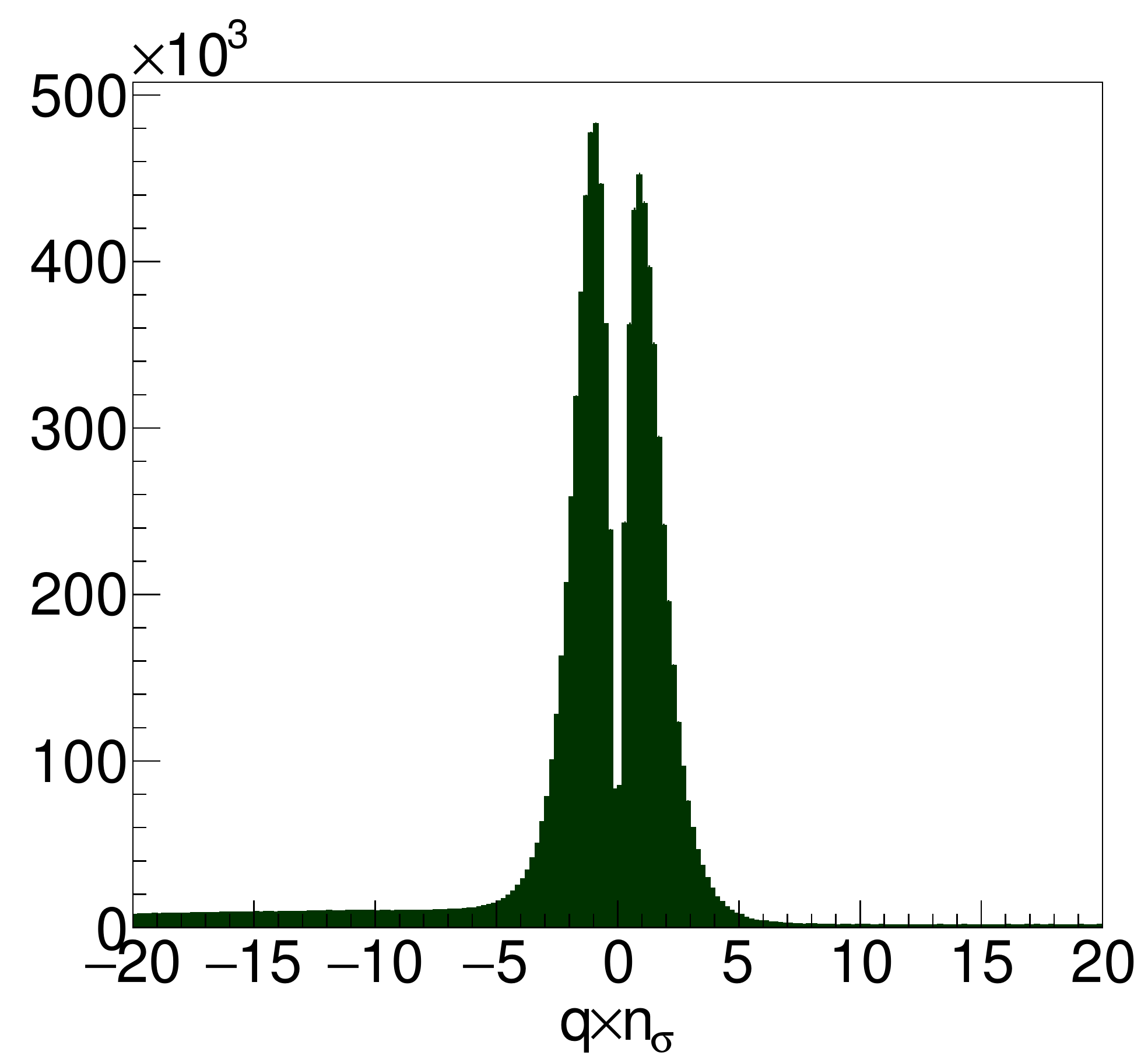}
\caption{Distribution of $q \times n_\sigma$ 
		for in-time clusters, where $q$ is the electric charge. 
		Positive tracks were weighted by $\approx 6$ to have visually the same level as negatives.
		}
\label{fig:nSigma_left}
\end{figure}

With this matching estimator $n_\sigma$ defined, we studied its distribution for all combinations of good tracks and clusters in the same detector half (shown in Fig.~\ref{fig:nSigma_left}).
It can be seen that good matching between tracks and clusters was achieved with small background. Moreover, the matching between the SVT and the ECal allowed the rejection of
about 9\% of negative tracks for which no match was found ($n_\sigma > 5$) and about 3\% of positive tracks.

\section{Summary}
With all 442 channels and all hardware components fully operational,
the HPS electromagnetic calorimeter performed well during the first runs of the experiment in 2015-2016.
Its primary goal of providing a fast trigger in a large background environment was achieved, with online cluster construction
and efficient cluster pair selection at a rate of up 30~kHz with only 10\% dead time.
In addition, detailed simulations and careful calibrations lead to energy and position resolutions
of about 4\% and 2 mm for 1~GeV electrons. A time resolution better than 1~ns for all hit energies above 0.2~GeV was obtained using only FADC pulse information.
The data presented in this paper was collected during engineering runs of 2015 and 2016; the physics runs are planned in the coming years at energies ranging from 1~GeV to 6.6~GeV to cover the planned search domain of the HPS experiment.

\section*{Acknowledgements}
The ECal could not have been designed and built without the participation of the skilled technical staff at IPNO (Orsay), INFN (Cagliari, Catania, Genova, Roma Tor Vergata and Torino) and Jefferson Lab.
The whole HPS collaboration participated in its commissioning and operation during the runs, and numerous discussions within the collaboration
helped to refine the analyses described in this article. The Jefferson Lab accelerator crew is thanked for delivering a well tuned beam extremely close
to our detection system. We also acknowledge the editing help of Norman A. Graf. This work  was supported by a grant from the French National Research Agency (ANR-13-JS05-0001), by a Sesame project HPS@JLab funded by the French region Ile-de-France, by the Italian Istituto Nazionale di Fisica Nucleare and by the U.S. Department of Energy, Office of Science,  Office of Nuclear Physics, under grant DE-FG02-96ER40960 and contract DE-AC05-06OR23177.



\section*{References}
\begin{thebibliography}{1}

\bibitem{REssig} R. Essig, J. A. Jaros, W. Wester et al.,  
	Planning the future of U.S. Particle Physics, The Intensity Frontier,  Dark Sectors and New, Light, Weakly Coupled Particles, 
	\href{http://arxiv.org/abs/1311.0029}{arXiv:1311.0029};  
	J. Alexander et al., 
	Dark Sectors 2016 Workshop: Community Report,
	\href{http://arxiv.org/abs/1608.08632}{arXiv:1608.08632}; 
	and references therein.
	
\bibitem{HPSTestRun} M. Battaglieri et al., The Heavy Photon Search Test Detector, Nucl. Instrum. Meth. A {\bf 777} (2014) 91-101.	

\bibitem{FX-thesis} F.-X. Girod, 
            Diffusion Compton profond\'ement virtuelle avec le d\'etecteur CLAS pour une \'etude des distributions de partons g\'en\'eralis\'ees,      
           \href{http://www.jlab.org/Hall-B/general/thesis/fxgirod.pdf}{PhD thesis}, 
	Universit\'e de Strasbourg (2006), and references therein.


\bibitem{APD} Hamamatsu \href{http://www.hamamatsu.com/jp/en/S8664-1010.html}{Si APD S8664-1010}.

\bibitem{preampli} E. Rauly and G. Charles, Current sensitive preamplifier used for HPS calorimeter, \href{https://misportal.jlab.org/mis/physics/hps_notes/viewFile.cfm/2016-001.pdf?documentId=17}{HPS-Note 2016-001}.

\bibitem{LEDbench} A. Celentano et al., Design and realization of a facility for the characterization of Silicon Avalanche PhotoDiodes, 
	JINST 9 (2014) T09002, \href{http://arxiv.org/abs/1504.01589}{arXiv:1504.01589}.

\bibitem{Andrea-Gabriel-Note} A. Celentano and G. Charles, Characterization of the ECal crystals light yield and amplification chain, 
	\href{https://misportal.jlab.org/mis/physics/hps_notes/viewFile.cfm/2014-002.pdf?documentId=2}{HPS-Note 2014-002}.

\bibitem{FADC} H. Dong et al., Integrated Tests of a High Speed VXS Switch Card and 250 MSPS Flash ADCs (2007).

\bibitem{EPICS} L.R. Dalesio et al.,
	\href{http://www.aps.anl.gov/epics/EpicsDocumentation/EpicsGeneral/EPICS_Architecture.pdf}
	        {EPICS Architecture} (1991). 

\bibitem{mya} C.J. Slominski, A MySQL Based EPICS Archiver,  
	\href{http://accelconf.web.cern.ch/AccelConf/icalepcs2009/papers/wep021.pdf}
	        {Proc. 12th Int. Conf. on Accelerator and Large Experimental Physics Control Systems}, 
	Kobe, Japan (2009).
 
\bibitem{Batarin} V.A. Batarin et al., 
                          Correlation of beam electron and LED signal losses under irradiation and long-term recovery of lead tungstate crystals, 
                          Nucl. Instrum. Meth. A {\bf 550} (2005) 543-550.
                          
\bibitem{Geant4} S. Agostinelli et al., Geant4—a simulation toolkit, Nucl. Instrum. Meth. A {\bf 506} (2003) 250-303.                          

\bibitem{Loo} T. C. Awes et al.,
		 A Simple Method of Shower Localization and Identification in Laterally Segmented Calorimeters, 
		Nucl. Instrum. Meth. A {\bf 311} (1992) 130.

\bibitem{SFraction} H. Szumila-Vance and M. Gar\c con ,	
                  HPS/ECal simulations: energy and position reconstruction for electrons, positrons and photons,
	\href{,https://misportal.jlab.org/mis/physics/hps_notes/viewFile.cfm/2014-001.pdf?documentId=1}{HPS-Note 2014-001}.
	
\bibitem{EnerCalNote} H. Szumila-Vance,	
                  HPS Ecal Energy Calibration for the Spring 2015 Engineering Run,
	\href{,https://misportal.jlab.org/mis/physics/hps_notes/viewFile.cfm/2016-002.pdf?documentId=18}{HPS-Note 2016-002}.	

\bibitem{NathansNote} N. Baltzell, ECal pulse fitting,
	\href{https://misportal.jlab.org/mis/physics/hps_notes/viewFile.cfm/2015-010.pdf?documentId=12}{HPS-Note 2015-010}.

\bibitem{TimeCalNote} H. Szumila-Vance, 
                     ECal Timing Calibration for the Spring 2015 Engineering Run,
                     \href{https://misportal.jlab.org/mis/physics/hps_notes/viewFile.cfm/2015-011.pdf?documentId=13}{HPS-Note 2015-011}



\end{thebibliography}
\end{document}